\documentclass[twocolumn]{aastex631}

\usepackage{graphicx}	
\usepackage{amsmath}	
\usepackage[T1]{fontenc}
\usepackage{natbib}
\usepackage{float}

\usepackage{ulem}

\usepackage{color}

\begin{document}
\title{Low angular momentum general relativistic magnetohydrodynamic accretion flow around rotating black holes with shocks}

\author[0000-0002-5730-0376]{Samik Mitra}
\affiliation{Indian Institute of Technology Guwahati, Guwahati 781039, Assam, India}

\correspondingauthor{sbdas@iitg.ac.in}
\email{m.samik@iitg.ac.in}

\author[0000-0003-4399-5047]{Santabrata Das}
\affiliation{Indian Institute of Technology Guwahati, Guwahati 781039, Assam, India}
\email{sbdas@iitg.ac.in}


\begin{abstract}

We investigate the global structure of general relativistic magneto-hydrodynamic (GRMHD) accretion flows around Kerr black holes containing shock waves, where the disk is threaded by radial and toroidal magnetic fields. We self-consistently solve the GRMHD equations that govern the flow motion inside the disk and for the first time to our knowledge, we obtain the shock-induced global GRMHD accretion solutions around weakly as well as rapidly rotating black holes for a set of fundamental flow parameters, such as energy ($\mathcal{E}$), angular momentum ($\mathcal{L}$), radial magnetic flux ($\Phi$), and iso-rotation parameter ($F$). We show that shock properties, namely shock radius ($r_{\rm sh}$), compression ratio ($R$) and shock strength ($\Psi$) strongly depends on $\mathcal{E}$, $\mathcal{L}$, $\Phi$, and $F$. We observe that shock in GRMHD flow continues to exist for wide range of the flow parameters, which allows us to identify the effective domain of parameter space in $\mathcal{L}-\mathcal{E}$ plane where shock solutions are feasible. Moreover, we examine the modification of the shock parameter space and find that it shifts towards the lower angular momentum values with increasing $\Phi$ and black hole spin ($a_{\rm k}$). Finally, we compute the critical radial magnetic flux ($\Phi^{\rm cri}$) that admits shocks in GRMHD flow and ascertain that $\Phi^{\rm cri}$ is higher (lower) for black hole of spin $a_{\rm k} = 0.99$ ($0.0$) and vice versa. 

\end{abstract}


\section{Introduction} \label{sec:intro}

Recent findings of large-scale magnetic fields surrounding supermassive black holes (SMBHs), as revealed by \cite{EHT_Mag2021}, indicates their potential influence on the accretion and ejection mechanisms. Earlier, theoretical proposition of \cite{Shakura-Sunyaev1973} suggests that angular momentum transport within an accretion disk could be facilitated by magneto-hydrodynamical (MHD) turbulence. Almost two decades later, the underlying physical mechanism for angular momentum transport is identified by \cite{Balbus-Hawley1991,Balbus-Hawley1998} with their seminal work on magneto-rotational instability (MRI). In reality, an accretion disk around black hole (BH) is expected to be threaded by large-scale magnetic fields, which are rooted either from the low-mass companion star or from the interstellar medium \citep{Kogan-Ruzmaikin1974,Kogan-Ruzmaikin1976,Kogan-Lovelace2011}. 

Meanwhile, numerous efforts were given to understand the nature and structure of the magnetic fields in an accretion flow around BH \cite[][and references therein]{Igumenshchev-etal2003,Shafee-etal2008,Begelman-Pringle2007,Mishra-etal2016,Mishra-etal2020}. Based on the geometry of the disk and plasma dynamics surrounding BHs, toroidal magnetic fields appear to be the simplest choice, as indicated by \cite{Oda-etal2007,Oda-etal2010,Oda-etal2012}, although the plunging region seems to be governed by the poloidal magnetic fields \cite[]{Hawley2001,Kato-etal2004}. \cite{DeVilliers-etal2003,Hirose-Krolik2004} performed 3D general relativistic magnetohydrodynamic (GRMHD) simulations and found that the plunging region is predominantly governed by poloidal magnetic fields although ordered toroidal magnetic fields play a significant role in regulating the dynamics of the inner regions of accretion disks. In addition, \cite{Avara-etal2016} numerically showed that in a radiatively efficient thin accretion disks, large-scale magnetic fields naturally accrete through the disk while enhancing the disk's radiative efficiency. \cite{Liska-etal2022} examined the behavior of two-temperature truncated disks using GRMHD simulations and found that large-scale net poloidal magnetic flux leads to the formation of a two-phase environment consisting of cold gas clumps moving within a hot magnetically dominated corona. Very recently, \cite{Manikantan-etal2024} performed GRMHD simulations where they initialized the disk with a toroidal magnetic field that dynamically evolved, giving rise to significant poloidal fields via magnetic dynamo process \cite[]{Jacquemin-Ide-etal2023}. It is worth mentioning that all these studies are model dependent and hence, exact configuration of the magnetic fields within the disk remains unresolved.

In magnetized accretion disk, flow starts accreting sub-sonically far from the BH. As flow moves towards a black hole, it gains radial velocity, reaching super-sonic speeds while crossing the event horizon. Hence, flow must change its sonic state to become transonic \cite[]{Fukue1987,Chakrabarti1989,Takahashi-etal1990,Takahashi-etal2002} at a radius commonly known as critical point. Depending on the flow parameters, namely energy, angular momentum and magnetic fields, flow may contain either single or multiple critical points \cite[]{Sarkar-Das2016,Sarkar-etal2018,Das-Sarkar2018,Mitra-etal2022}. Note that multi-transonic flows often exhibit discontinuous shock transitions \cite[]{Fukue1987,Chakrabarti1989,Das-etal2001,Takahashi-etal2002}. During advection, rotating matter experiences centrifugal repulsion, leading to the accumulation of matter in the vicinity of the black hole. This forms a `virtual' barrier around the black hole triggering the shock transition when possible. Indeed, accretion solutions containing shocks are thermodynamically preferred due to their high entropy content \cite[]{Becker-Kazanas2001}, which facilitates in explaining spectro-temporal signatures of black hole X-ray binaries \cite{Chakrabarti-Titurchuk1995,Mandal-Chakrabarti2005,Nandi-etal2012,Iyer-etal2015,Das-etal2021,Majumder-etal2022,Nandi-etal2024}. Realizing the astrophysical significance shock-induced accretion solutions are studied both in hydrodynamics \cite[]{Fukue1987,Chakrabarti1989,Yang-Kafatos1995,Ryu-etal1997,Lu-etal1999,Gu-Lu2001,Becker-Kazanas2001,Chakrabarti-Das2004,Das2007,Becker-etal2008,Das-etal2009,Kumar-etal2013,Das-etal2014,Sukova-Janiuk2015,Sukova-etal2017,Aktar-etal2017,Kim-etal2019,Dihingia-etal2019,Sen-etal2022} as well as magneto-hydrodynamic \cite[]{Takahashi-etal2006,Fukumura-etal2007,Sarkar-Das2015,Sarkar-Das2016,Fukumura-etal2016,Das-Sarkar2018} scenarios. However, efforts are pending in investigating the accretion dynamics involving shocks in GRMHD flow around rotating black holes.

Motivating with this, we study the MHD accretion flows around Kerr BHs of spin $a_{\rm k}$ under the general relativistic frame work. The GRMHD flow under consideration is characterized by means of radial magnetic flux ($\Phi$) and iso-rotation parameter ($F$) \cite[]{McKinney-Gammie2004} in addition to flow energy ($\mathcal{E}$) and angular momentum ($\mathcal{L}$). With this, we obtain the shock-induced global accretion solutions adopting the relativistic equation of state \cite[REoS;][]{Chattopadhyay-Ryu2009} for the first time to the best of our knowledge. We find that shocked solutions exist around both weakly rotating ($a_{\rm k} \rightarrow0$) as well rapidly rotating ($a_{\rm k}=0.99$) BHs. We examine the shock properties, namely shock location ($r_{\rm sh}$), compression ratio ($R$) and shock strength ($\Psi$) and find that these quantities strongly depends on the model parameters $\mathcal{E}$, $\mathcal{L}$, $\Phi$, and $F$. Moreover, we observe that shocks in GRMHD flow continuous to form for wide range of model parameters. Hence, we separate the parameter space in $\mathcal{L}-\mathcal{E}$ plane to identify regions where shocked GRMHD solutions are feasible, and also examine its modifications with $\Phi$ and $a_{\rm k}$. We further calculate the critical radial magnetic flux ($\Phi^{\rm cri}$) that admits shock in GRMHD flow and find that $\Phi^{\rm cri}$ is higher for rapidly rotating BHs compared to non-rotating black hole. Finally, we indicate that the GRMHD shocked accretion flows seem to fail in reaching the MAD limit \cite[]{Igumenshchev-etal2003, Narayan-etal2003, Sadowski2016b}. 

The paper is organized as follows. In Section {\ref{EQN}}, we describe the GRMHD equations and the underlying model assumptions. In Section \ref{SHOCK}, we discuss GRMHD shock solutions. In Section \ref{RESULTS}, we discuss the obtained results. Finally, we summarize the overall findings in Section \ref{Conclusion}.

\section{Magnetized flow: Formalism and underlying assumptions} \label{EQN}

In this paper, we aim to study the magnetized hot accretion flows around a stationary, axisymmetric rotating BH. In Boyer-Lindquist coordinates, the line element of rotating BH space-time takes the following form \citep{Kerr1963}:
\begin{eqnarray}\label{rotmetric}
ds^2 & = & - \left( 1- \frac{2r}{\Sigma} \right) dt^2  - \frac{4a_{\rm k}r}{\Sigma  } \sin^2 \theta dt \; d\phi +
\frac{\Sigma}{\Delta}dr^2  \nonumber
\\ & &+ \Sigma d \theta^2+ \left[r^2+ a_{\rm k}^2 +
\frac{2 r a_{\rm k}^2 }{\Sigma} \sin^2 \theta
\right] \sin^2 \theta d\phi^2,
\end{eqnarray}
where $\Sigma = r^2 + a_{\rm k}^2 \cos^2\theta$, $\Delta = r^2 + a_{\rm k}^2 - 2r$ and $a_{\rm k}$ is the BH spin. In this work, we express length $r$ and time $t$ in terms of $r_{\rm g}$ and $r_{\rm g}/c$, $r_{\rm g}=GM_{\rm BH}/c^2$ being gravitational radius, where $G$ is the gravitational constant, $c$ is the speed of light, and $M_{\rm BH}$ is the mass of BH. With this unit system, we write the governing GRMHD equations \citep[and references therein]{Lichnerowicz1970, Anile1990} as,
\begin{equation}
\nabla_\mu (\rho u^\mu) = 0; \hspace{0.5cm} \nabla_\mu T^{\mu\nu}=0; \hspace{0.5cm} \nabla_\mu {}^* F^{\mu\nu}=0,
\end{equation} where $\rho$ is the mass density, $u^\mu$ is the four velocity, $T^{\mu \nu}$ is the energy-momentum tensor, and ${}^*F^{\mu\nu}$ is the dual of Faraday's electromagnetic tensor. We consider the accretion flow with infinite conductivity that allows the magnetic field lines to remain frozen into the accreting plasma according to the ideal GRMHD conditions, \textit{i.e.,} $u_\mu b^\mu =0$. In a magnetized flow, the energy-momentum tensor is given by \citep{Abramowicz-Fragile2013},
\begin{equation}
\begin{aligned}
T^{\mu}_\nu &=(T^\mu_\nu)_{\rm Fluid} + (T^\mu_\nu)_{\rm Maxwell}\\
&= \rho \bigg( \frac{e+p_{\rm gas} + b^2}{\rho}\bigg) u^\mu u_\nu + \delta^{\mu}_{\nu} \bigg(p_{\rm gas} + \frac{b^2}{2}\bigg) - b^\mu b_\nu.
\end{aligned}
\end{equation}
Here, $p_{\rm gas}$ is the gas pressure, $e$ is the internal energy, $b^\mu$ refers to the four magnetic fields in the comoving frame, and $b^2=b_\mu b^\mu$. Note that, $\delta^\mu_\nu = g^{\mu\alpha} g_{\alpha\nu}$ is the contraction of the covariant and contravariant metric components.

\subsection{Conserved quantities in GRMHD flows}

We consider the convergent flow ($u^r < 0$) to be confined in the disk mid-plane, \textit{i.e.,} $\theta=\pi/2$ and hence, the polar component of the four-velocity tends to become zero as $u^\theta \sim 0$. Furthermore, we choose $b^\theta=0$, making the radial ($b^r$) and toroidal ($b^\phi$) field components independent. With this, we solve the radial behavior of the advective, axisymmetric ($\partial_\phi \rightarrow 0$) flow in the steady-state ($\partial_t \rightarrow 0$).  

From the particle number conservation equation, we get
\begin{equation}
	\sqrt{-g} \rho u^r = {\rm constant} = C_{\mathcal{M}},
\end{equation}
where $C_{\mathcal{M}}$ is a measure of the mass flux and for Kerr BH, the determinant $\sqrt{-g}=r^2$ in $\theta=\pi/2$ limit. Being stationary and axisymmetric, the Kerr metric is associated with two Killing vector fields. As the fluid is assumed to obey the symmetries of the chosen space-time, the energy-momentum conservation takes the form $\nabla_\mu(T^{\mu}_\nu \xi^\nu)=0$, where $\xi^\nu$ refers to the generic killing vectors. Accordingly, we obtain the globally conserved specific energy flux ($\mathcal{E}$) for $\nu=t$ as

\begin{equation}
	-\frac{\sqrt{-g} T^r_t}{C_{\mathcal{M}}} = \mathcal{E},
\end{equation}
and the conserved specific angular momentum flux ($\mathcal{L}$) is obtained for $\nu=\phi$ as
\begin{equation}
	\frac{\sqrt{-g} T^r_\phi}{C_{\mathcal{M}}} = \mathcal{L}.
\end{equation}

Additionally, the no-monopole constrain \cite[]{Porth-etal2019} implies,
\begin{equation}
-\sqrt{-g}{}\, {}^*F^{rt}={\rm const}=\Phi,
\end{equation}
where $ {}^*F^{rt} =u^t b^r - u^r b^t$. The $\phi$-component of source-free Maxwell's equation implies \cite[]{McKinney-Gammie2004},
\begin{equation}
\sqrt{-g}{}{}^*F^{r\phi}={\rm const}=F,
\end{equation}
where ${}^*F^{r\phi}=u^r b^\phi - u^\phi b^r$. It is noteworthy that equation (8) is commonly known as the relativistic iso-rotation law. Finally, we obtain the $r$-component of the Navier-Stokes equation by projecting the energy-momentum conservation equation along the radial direction in the fluid frame \cite[]{Mitra-etal2022}, which is given by,
\begin{equation}
\begin{aligned}
       &\gamma^r_\mu \nabla_\nu T^{\mu\nu}=0,\\
       &(g^{r\nu}+u^ru^\nu) \nabla_\nu p_{\rm tot} + \rho h_{\rm tot} u^\nu \nabla_\nu u^\alpha - \nabla_\nu (b^r b^\nu) \\ &- u^r u_\mu \nabla_\nu (b^\mu b^\nu)=0,
\end{aligned}
\end{equation}
where $\gamma^r_\mu ~(= \delta^r_\mu + u^r u_\mu)$ is the projection operator, $p_{\rm tot}~(=p_{\rm gas} + p_{\rm mag})$ is the total pressure, $p_{\rm mag}~(=b^2/2$) is the magnetic pressure, and $h_{\rm tot}~[= (e+p_{\rm gas})/\rho + b^2/\rho]$ is the total enthalpy.

Following \cite{Riffert-Herold1995, Peitz-Appl1997}, we calculate the local half-thickness ($H$) of the magnetized disk considering hydrostatic equilibrium in the vertical direction, which is given by,
\begin{equation}
      H^2 = \frac{p_{\rm gas} r^3}{\rho \mathcal F}, \hspace{0.5cm} \mathcal{F}=\gamma_\phi^2 \frac{(r^2+a_{\rm k}^2)^2 + 2 \Delta a_{\rm k}^2}{(r^2+a_{\rm k}^2)^2 - 2 \Delta a_{\rm k}^2},
\end{equation}
where $\gamma_\phi$ $(=1/\sqrt{1-v_\phi^2})$ is the Lorentz factor. We define the specific angular momentum of the flow as $\lambda$ $(=-u_\phi/u_t)$ and the angular velocity of the flow is given by $\Omega$ $(=u^\phi/u^t)$ \cite[]{Dihingia-etal2018b, Mitra-etal2022}. We follow \cite{Lu1985} to describe the three components of fluid velocities in the corotating frame as $v_\phi^2 = u^\phi u_\phi/(-u^t u_t)$, $v_r^2 = u^r u_r/(-u^t u_t)$, and $v_\theta^2 = u^\theta u_\theta/(-u^t u_t)$, where $v_\theta=0$ as $u^\theta \sim 0$ in the disk mid-plane. Upon integrating equation (4), we obtain the globally conserved mass-accretion rate in the comoving frame, which is given by,
\begin{equation}
	\dot{M} = -4\pi \rho v \gamma_v H \sqrt{\Delta},
\end{equation}
where $v$ $(= \gamma_\phi v_r)$ is the flow velocity and $\gamma_v=1/\sqrt{1-v^2}$. In this work, we express the accretion rate as $\dot{m}=\dot{M}/\dot{M}_{\rm Edd}$, where $\dot{M}_{\rm Edd}$ is the Eddington accretion rate ($\dot{M}_{\rm Edd} = 1.4 \times 10^{18} \big(\frac{M_{\rm BH}}{M_\odot}\big)$ g s$^{-1}$, $M_\odot$ being the solar mass). In this work, we choose ${\dot m}=0.001$. Using ideal MHD condition $u_\mu b^\mu=0$ and equations (7-8), we express $b^{r,\phi}$ in terms of $\Phi$ and $F$ as
\begin{equation}
b^r = - \frac{\gamma_\phi^2 (\Phi + F \lambda)}{u^t \mathcal{A}},\;\;\;
b^\phi = \frac{F v^2 - \gamma_\phi^2 (F + \Phi \Omega)}{u^r \mathcal{A}},
\end{equation}
where $\mathcal{A}=r^2 (v^2 - 1)$. Adopting the transformations in equation (12), we analyze the magnetized accretion flow in terms of the global constants $\Phi$ and $F$, respectively.

\subsection{Equation of state}
In order to close the dynamical equations [\textit{i.e.,} equations (5-9) and equation (11)], we use the relativistic equation of state (REoS; \citealp[]{Chattopadhyay-Ryu2009}), which is given by,
\begin{equation}
	e = \frac{\rho f}{\big(1+\frac{m_p}{m_e}\big)},
\end{equation}
where $m_e$ and $m_p$ are the masses of electrons and ions. The quantity $f$ is expressed in terms of dimensionless temperature ($\Theta = k_{\rm B} T/m_e c^2$, $k_{\rm B}$ is the Boltzmann constant) as
\begin{equation}
f = \Bigl\{1+ \Theta \bigg(\frac{9\Theta+3}{3\Theta+2}\bigg)\Bigl\} + \Bigl\{\frac{m_p}{m_e}+\Theta \bigg(\frac{9\Theta m_e+3m_p}{3\Theta m_e+2m_p}\bigg)\Bigl\}.
\end{equation}
With this, we define the polytropic index as $N=(1/2) (df/d\Theta)$ and adiabatic index as $\Gamma = 1 + 1/N$. Notably, the characteristic wave speeds for magnetized flows are associated with the Alfv\'en and magneto-sonic waves, respectively. Following \cite{Gammie-etal2003}, we define the Alfv\'en speed as $C_a^2 = b_\mu b^\mu/(\rho h_{\rm tot})$, and the fast-magnetosonic speed as $C_{\rm f}^2= C_{\rm s}^2 + C_a^2 - C_{\rm s}^2 C_a^2$, where the relativistic sound speed is given by $C^2_{\rm s}=\Gamma p_{\rm gas}/(e + p_{\rm gas})$. Moreover, we define the magnetosonic Mach number as $M=v/C_{\rm f}$. 

\subsection{Critical point analysis} \label{Method}

We combine equations (5-9, 11) and obtain three coupled non-linear differential equations as,
\begin{itemize}
\item[(a)] the radial momentum equation:
\begin{equation}
\begin{aligned}
R_0 + R_v \frac{dv}{dr} + R_\Theta \frac{d\Theta}{dr} + R_\lambda \frac{d\lambda}{dr} &=0,
\end{aligned}
\end{equation}
\item[(b)] the azimuthal momentum equation:
\begin{equation}
\begin{aligned}
\mathcal{L}_0 + \mathcal{L}_v \frac{dv}{dr} + \mathcal{L}_\Theta \frac{d\Theta}{dr} + \mathcal{L}_\lambda \frac{d\lambda}{dr} &=0,
\end{aligned}
\end{equation}
\item[(c)] the energy equation:
\begin{equation}
\begin{aligned}
\mathcal{E}_0 + \mathcal{E}_v \frac{dv}{dr} + \mathcal{E}_\Theta \frac{d\Theta}{dr} + \mathcal{E}_\lambda \frac{d\lambda}{dr} &=0.
\end{aligned}
\end{equation}
\end{itemize}
The coefficients ($R_j, \mathcal{L}_j, \mathcal{E}_j$; $j\rightarrow 0, v, \Theta$, and $\lambda$) in equations (15-17) are given in Appendix-A.

Using equations (15-17), we obtain the wind equation of the GRMHD flow after some simple algebra as,
\begin{equation}
\frac{dv}{dr} = \frac{\mathcal{N}(r,a_{\rm k},v,\Theta,\lambda,\Phi,F)}{\mathcal{D}(r,a_{\rm k},v,\Theta,\lambda,\Phi,F)},
\end{equation}
where the explicit expression of numerator $\mathcal{N}$ and denominator $\mathcal{D}$ are given in Appendix-B. Moreover, we express the gradients of the angular momentum ($\lambda$) and temperature ($\Theta$) in terms of $dv/dr$ as  
\begin{equation}
\frac{d\lambda}{dr} = \frac{\mathcal{L}_{\Theta} \mathcal{E}_{0} - \mathcal{L}_{0} \mathcal{E}_{\Theta}}{\mathcal{L}_{\lambda} \mathcal{E}_{\Theta} - \mathcal{L}_{\Theta} \mathcal{E}_{\lambda}} + \frac{(\mathcal{L}_{\Theta} \mathcal{E}_{v} - \mathcal{L}_{v} \mathcal{E}_{\Theta})}{\mathcal{L}_{\lambda} \mathcal{E}_{\Theta} - \mathcal{L}_{\Theta} \mathcal{E}_{\lambda}} \frac{dv}{dr},
\end{equation}
and 
\begin{equation}
\frac{d\Theta}{dr} = \frac{\mathcal{L}_{\lambda} \mathcal{E}_{0} - \mathcal{L}_{0} \mathcal{E}_{\lambda}}{\mathcal{L}_{\Theta} \mathcal{E}_{\lambda} - \mathcal{L}_{\lambda} \mathcal{E}_{\Theta}} + \frac{(\mathcal{L}_{\lambda} \mathcal{E}_{v} - \mathcal{L}_{v} \mathcal{E}_{\lambda})}{\mathcal{L}_{\Theta} \mathcal{E}_{\lambda} - \mathcal{L}_{\lambda} \mathcal{E}_{\Theta}} \frac{dv}{dr}.
\end{equation}
In order to obtain GRMHD accretion solutions around rotating BH, we simultaneously solve equations (18-20) for a set of model parameters, namely $\mathcal{E}$, $\mathcal{L}$, $\Phi$, $F$ and $a_{\rm k}$, respectively. 

Usually, the accreting matter begins its journey from the outer edge ($r_{\rm edge}$) of the disk with subsonic radial velocity ($v << 1$) and descends into the BH super-sonically ($v\sim 1$) to fulfill the inner boundary conditions imposed by the horizon. Therefore, the flow must become trans-magnetosonic at least once, if not more, while passing through critical point ($r_c$).  At the critical point, the wind equation (equation 18) takes an indeterminate form as $(dv/dr)|_{r_{\rm c}} = 0/0$ that yields the critical point conditions $\mathcal{N}=\mathcal{D}=0$. However, in reality, a convergent flow always remains smooth along the streamlines even while passing through $r_{\rm c}$. Hence, the velocity gradient must be real and finite everywhere. We, therefore, imply l'H$\hat{{\rm o}}$pital rule in equation (18) to evaluate the velocity gradient at $r_{\rm c}$. Accordingly, we obtain two unique values of $(dv/dr)|_{r_{\rm c}}$; one of them relates to accretion, and the other one is for wind. When both $(dv/dr)|_{r_{\rm c}}$ values are real and of opposite in sign, saddle type critical point is formed \cite[]{Das2007,Das-etal2022,Mitra-etal2022,Mitra-etal2023}. Such points have special significance, as trans-magnetosonic solutions can only pass through these points before entering into BH. Depending on the model parameters, when critical point forms close to the horizon, it is named as inner ($r_{\rm in}$) critical point, whereas the outer one ($r_{\rm out}$) is formed far away from the horizon. Notably, the GRMHD accretion flow around BHs often possesses multiple critical points (MCP) depending on the model parameters $(a_{\rm k},\mathcal{E},\mathcal{L},\Phi, F)$. Such GRMHD flows are potentially viable to harbour shock waves \cite[][and references therein]{Fukue1987, Chakrabarti1989, Das-Chakrabarti2004, Das2007, Dihingia-etal2019, Das-etal2022}. Accordingly, in the subsequent sections, we study the shock-induced magnetized accretion flows around rotating BH adopting the general relativistic framework.

\section{Shock induced GRMHD accretion  flow}
\label{SHOCK}

Depending on the model parameters ($a_{\rm k},\mathcal{E},\mathcal{L},\Phi,F$), GRMHD flow becomes supersonic after passing through the outer critical point ($r_{\rm out}$), and continues to proceed towards the BH. Meanwhile, flow starts experiencing centrifugal repulsion, which momentarily slows down the accreting matter. Because of this, matter accumulates in the vicinity of the BH, and a barrier is formed in the form of an effective boundary layer around BH. Such a centrifugal barrier cannot hold the accumulation of matter indefinitely and beyond a critical limit, it eventually triggers the discontinuous transition of the flow variables in the form of shock waves \cite[]{Fukue1987, Chakrabarti1989, Frank-etal2002,Takahashi-etal2002, Das-Chakrabarti2007,Fukumura-etal2007,Sarkar-Das2016, Sarkar-etal2018, Dihingia-etal2019, Dihingia-etal2020}. Considering this, we describe the shock conditions for MHD flow in the GR framework below, which eventually enables us to study shock properties, namely shock location ($r_{\rm sh}$), shock compression ratio ($R$), and shock strength ($\Psi$).

In order to execute discontinuous shock transition, magnetized accretion flow must satisfy the general relativistic shock conditions \cite[]{Lichnerowicz1970, Appl-Camenzind1988, Takahashi-etal2006, Fukumura-etal2007}, which are given by, 
\begin{itemize}
\item [(a)] Mass flux conservation, $\bigg[\rho u^r\bigg]=0,$
\item [(b)] Energy flux conservation, $\bigg[\frac{T^{rt}}{\rho u^r}\bigg]=0,$
\item [(c)] Angular momentum flux conservation, $\bigg[\frac{T^{r\phi}}{\rho u^r}\bigg]=0,$ 
\item [(d)] Radial magnetic flux conservation, $\bigg[{}^*F^{r t}\bigg]=0$,
\item [(e)] Iso-rotation conservation, $\bigg[{}^*F^{r\phi}\bigg]=0$,
\item [(f)] Pressure balance condition, $\bigg[\frac{T^{rr}}{\rho u^r}\bigg]=0.$
\end{itemize}
Here, the square bracket `$[\,\,]$' denotes the difference of a quantity across the shock front. Using these conditions, we obtain shock-induced global GRMHD accretion solution around rotating BH. Note that in this work, we assume the shocks to be thin and non-dissipative in nature for simplicity. 

Across the shock front ($r_{\rm sh}$), supersonic pre-shock flow jumps into the subsonic branch, resulting a hot and dense post-shock flow (equivalently post-shock corona, hereafter PSC, \cite{Aktar-etal2015}). This happens because the kinetic energy of pre-shock flow is converted into thermal energy and post-shock flow becomes compressed due to shock compression. This yields the PSC to act as a perfect reservoir of hot electrons which eventually intercepts the soft photons from the cooler pre-shock flow and reprocesses them to produce high-energy radiations via inverse Comptonization \cite[]{Chakrabarti-Titurchuk1995}. After the shock transition, the subsonic flow gradually gains its radial velocity and ultimately enters into BH supersonically after crossing the inner critical point ($r_{\rm in}$).

\section{Results} \label{RESULTS}

We investigate the dynamical structure of shock-induced trans-magnetosonic accretion solutions around BH of spin $a_{\rm k}$ for a set of model parameters, namely $\mathcal{E}$, $\mathcal{L}$, $\Phi$ and $F$, respectively. In doing so, we examine the effects of the radial magnetic flux ($\Phi$) and iso-rotation parameter $(F)$ on the GRMHD solutions. Given the diminutive nature of the dimensionless radial magnetic flux and iso-rotation parameter, we denote them as $\Phi = \Phi_{13} \times 10^{-13}$ and $F = F_{15} \times 10^{-15}$, maintaining the notation $\Phi_{13}$ and $F_{15}$ to signify magnetic flux values. Moreover, in this work, we choose $M_{\rm BH}=10M_\odot$ and $\dot{m}=0.001$ as fiducial values unless stated otherwise.  

\subsection{Shock-induced global GRMHD accretion solutions}

\begin{figure*}
    \includegraphics[width=\textwidth]{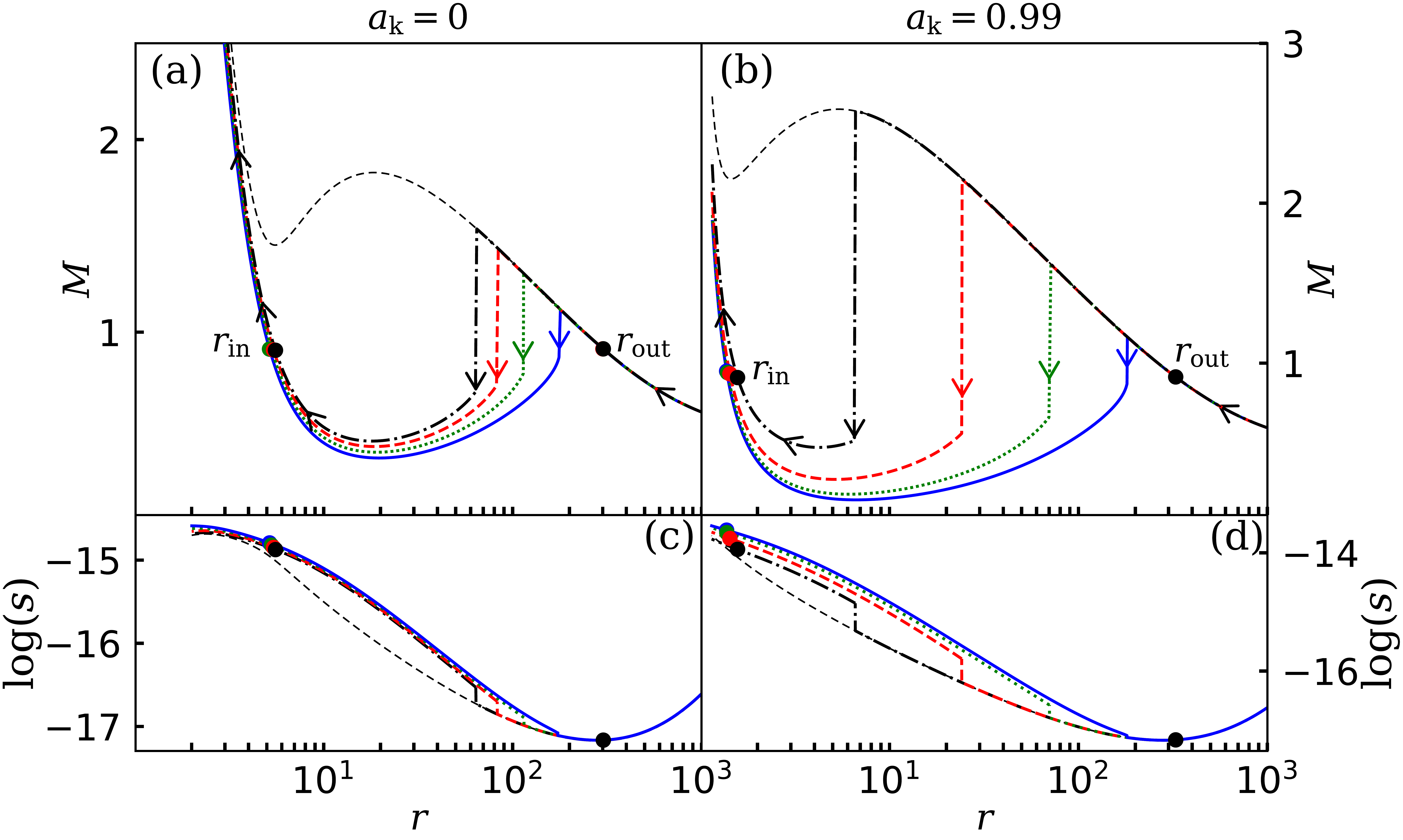}
    \caption{Plot of fast-magnetosonic Mach number ($M=v/C_{\rm f}$) with radial coordinate ($r$) for shock-induced accretion solutions around BHs. The chosen model parameters are $(\mathcal{E},F_{15})$=($1.001,7.5$). (a) Dot-dashed, dashed, dotted and solid curves denote the accretion solutions for $\Phi_{13}=0.0$, $5.0$, $7.5$ and $9.4$, respectively, where $\mathcal{L}=3.15$ and $a_{\rm k}=0.0$. (b) Dot-dashed, dashed, dotted and solid curves denote the accretion solutions for $\Phi_{13}=0.0$, $5.0$, $6.5$ and $7.12$, respectively, where $\mathcal{L}=1.95$ and $a_{\rm k}=0.99$. In both panels, vertical arrows indicate the shock transitions at radii ($r_{\rm sh}$) and filled circles denote the critical points ($r_{\rm in}$ and $r_{\rm out}$). Arrows indicate the overall direction of the flow motion towards BH. Entropy density ($s$) associated with the solutions presented in (a) and (b) are presented in (c) and (d), respectively. See the text for the details.}
    \label{fig:1}
\end{figure*}

To begin with, we consider an advective flow that accretes towards a non-rotating black hole starting from the outer edge of the disk at $r_{\rm edge}=1000$. The flow is characterized with the model parameters as $\mathcal{E} = 1.001$, $\mathcal{L}=3.15$, $F_{15}=7.5$ and $a_{\rm k}=0$. The obtained results are plotted in Fig. \ref{fig:1}a, where fast-magnetosonic Mach number ($M=v/C_{\rm f}$) is plotted with radial coordinate ($r$) for accretion solutions containing shock waves. Here, we observe that for $\Phi_{13}=0.0$, subsonic flow changes its sonic state after crossing the outer critical point at $r_{\rm out}=302.235$ to become supersonic. While the supersonic flow can smoothly enter into the black hole (thin dashed curve), it undergoes a discontinuous shock transition at $r_{\rm sh}=63.881$ (dot-dashed vertical arrow), as the entropy content of the post-shock branch is higher compared to the pre-shock flow \cite[]{Das-etal2001}. This is not surprising as it happens in accordance with the second law of thermodynamics due to the fact that shocked solution is thermodynamically preferred \cite[]{Becker-Kazanas2001} over the shock free solution. After the shock, the subsonic flow gradually gains radial velocity as it moves inward and crosses the inner critical point at $r_{\rm in}=5.556$ before entering the black hole supersonically. This result is presented using dot-dashed (blue) curve. Next, we increase the radial magnetic flux as $\Phi_{13}=5.0$, keeping the other model parameters fixed, that increases of magnetic pressure leading to the rise of total pressure ($p_{\rm tot}$). This eventually pushes the shock front outwards and shock settles down to a larger radius at $r_{\rm sh} = 83.088$. This result is shown using dashed (red) curves and dashed vertical arrow denotes the location of shock transition. For further increase of radial magnetic flux as $\Phi_{13}=7.5$, the shock transition happens at $r_{\rm sh}=113.975$ and shock-induced GRMHD solution is depicted using dotted (green) curve. Needless to mention that an indefinite increase of $\Phi_{13}$ is not possible, because beyond a critical limit of radial magnetic flux $\Phi_{\rm 13}=9.4$, shock ceases to exist as the shock conditions (Sec. \ref{SHOCK}) are not satisfied. In the figure, we denote this solution using solid (blue) curve. Furthermore, following  \cite{Das-etal2009,Porth-etal2017,Mitra-etal2022}, we compute the specific entropy function ($s\propto p_{\rm tot}/\rho^{\Gamma -1}$) corresponding to the shocked accretion solutions delineated in Fig. \ref{fig:1}a and plot it as function of $r$ in panel Fig. \ref{fig:1}c. We observe that in all cases, $s$ jumps to higher value at the shock radius ($r_{\rm sh}$), which evidently confirms that shocked accretion solutions possess higher entropy than the shock free solution.

We continue to examine the effect of $\Phi_{13}$ on the flow solutions for rapidly rotating BH as well. Towards this, we choose $a_{\rm k}=0.99$ and keep the remaining model parameters unchanged ($i.e.$, $\mathcal{E}=1.001$ and $F_{15}=7.5$) as in Fig. \ref{fig:1}a, except $\mathcal{L}=1.95$ to obtain the shock-induced magnetized accretion solutions. The obtained results are shown in Fig. \ref{fig:1}b. Note that we use lower $\mathcal{L}$ value for higher $a_{\rm k}$ simply because low angular momentum flow ($\mathcal{L}$) can only sustain shocks around rapidly rotating BHs \cite[]{Dihingia-etal2019,Sen-etal2022}. We observe that for $\Phi_{13}=0$, shock transition happens at a relatively smaller radius at $r_{\rm sh}=6.525$ (denoted by dot-dashed vertical arrow) compared to the non-spinning case. This seems to happen as the lower $\mathcal{L}$ resulted weak centrifugal repulsion and hence, shock front moves further inward. However, when $\Phi_{13}$ is increased to $5.0$, $6.5$ and $7.0$, the shock front moves outward as expected, and we obtain $r_{\rm sh}=24.25$  (dashed in red), $70.07$ (dotted in green), and $180.74$ (solid in blue), respectively. Note that beyond $\Phi_{13}=7.12$, shock conditions are not favourable and hence, shock transition ceases to exist for the chosen set of model parameters. We further notice that Mach number ($M$) of the relativistic magnetized flow generally remain restricted as $M \ge 3$ around rapidly spinning BH mainly due to the rapid increase in sound speed close to the horizon. This happens because of the frame-dragging effect \cite[]{Fukumura-Kazanas2007} around a rotating BH. Here, the rotation of BH compels the matter to corotate with the BH along $\phi$-direction before getting trapped by the strong gravitational pull. This essentially heats up the inner disk, and hence, sound speed increases. Indeed, a similar finding is observed for general relativistic hydrodynamic flows as well \cite[]{Dihingia-etal2018b}. Next, in Fig. \ref{fig:1}d, we present the plot of specific entropy function ($s$) with $r$ for solutions presented in Fig. \ref{fig:1}b, and find that for all instances, $s$ undergoes a significant increase at $r_{\rm sh}$. This provides a clear evidence that shocked accretion solutions possess higher entropy compared to the solution without a shock.

\subsection{Flow variables of shocked-induced GRMHD accretion solutions}

\begin{figure}
    \centering
    \includegraphics[width=\columnwidth]{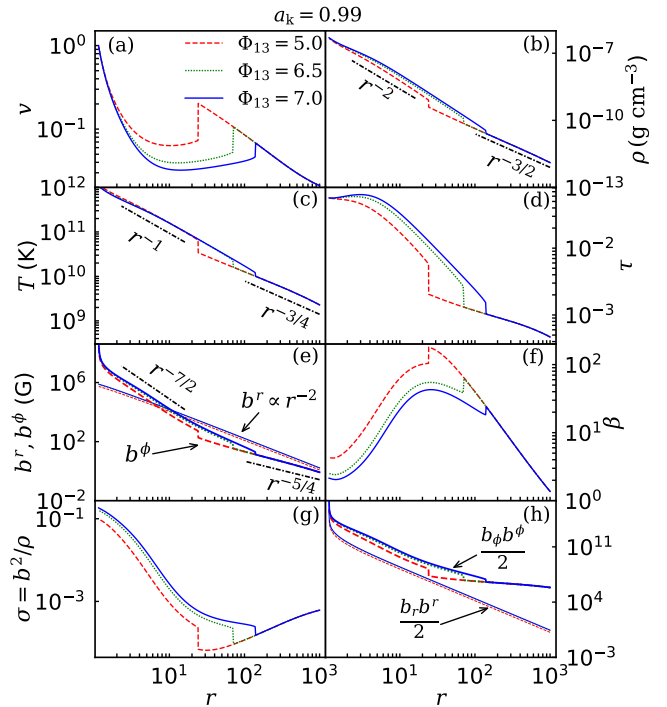}
    \caption{Radial variation of the primitive flow variables corresponding to shock-induced GRMHD accretion solutions presented in Fig. \ref{fig:1}b. In panels (a-h), profiles of radial velocity ($v$), density ($\rho$), temperature ($T$), scattering optical depth ($\tau$), radial ($b^r$) and toroidal ($b^\phi$) magnetic fields, plasma-$\beta$, magnetization ($\sigma=b^2/\rho$), and radial and toroidal magnetic pressures are plotted for different $\Phi_{13}$. Dashed (red), dotted (green) and solid (blue) curves denote results for $\Phi_{13}=5.0$, $6.5$ and $7.12$, respectively. In panels (b), (c) and (e), dot-dashed lines represent best-fit power law profiles of pre- and post-shock flow variables. See the text for the details.}
    \label{fig:vary_phi_akp99}
\end{figure}

In Fig. \ref{fig:vary_phi_akp99}, we investigate the behavior of various flow variables corresponding to the shocked GRMHD solutions depicted in Fig. \ref{fig:1}. In Fig. \ref{fig:vary_phi_akp99}a, we present the radial velocity ($v$) variation as a function of $r$ around a rapidly rotating BH of spin $a_{\rm k}=0.99$. We observe that for a set of model parameters $\mathcal{E}=1.001$, $\mathcal{L}=1.95$, and $F_{15}=7.5$, flow velocity monotonically increases in the pre-shock region and discontinuously drops down to the subsonic branch at the shock radius $r_{\rm sh}$. After the shock transition, flow momentarily slows down, although it gradually picks up radial velocity and ultimately enters into the BH with a velocity comparable to speed of light ($c$). Results plotted using dashed (red), dotted (green) and solid (blue) are obtained for $\Phi_{13} = 5.0$, $6.5$ and $7.12$, respectively, which are marked in the figure, and vertical lines denote the shock transition radii. In Fig. \ref{fig:vary_phi_akp99}b, we show the variation of mass density ($\rho$) with $r$, where sudden increase in $\rho$ is observed just after the shock transition for all $\Phi_{13}$ values. This happens because the radial velocity decreases at shock and hence, $\rho$ increases to higher value in order to preserve the conservation of mass-flux across the shock front. We observe that post-shock density profile follows a steeper power-law as $\rho \propto r^{-2}$, whereas the pre-shock density follows $\rho \propto r^{-3/2}$. Note that the pre-shock density profile exactly matches with the self-similar solutions for a pure inflow model in absence of outflows \cite[]{Narayan-Yi1995}. In Fig. \ref{fig:vary_phi_akp99}c, we depict the variation of flow temperature ($T$) with $r$. During the shock transition, the supersonic flow jumps into the subsonic branch and loses most of its kinetic energy that results in a hot post-shock flow. We notice that the temperature profile follows $T \propto r^{-1}$, which is commonly observed in Radiatively Inefficient Accretion Flow (RIAF) simulations \cite[]{Olivares-etal2023}. However, in the pre-shock regime, flow maintains a relatively shallower profile as $T \propto r^{-3/4}$. Due to shock compression, the hot and dense post-shock flow becomes puffed up resulting in an effective boundary layer (PSC) surrounding the BH. The presence of such coronal structure significantly affects the emergent radiations from the disk \cite[and references therein]{Chakrabarti-Titurchuk1995,Poutanen1998,Nandi-etal2012}. Keeping this in mind, in Fig. \ref{fig:vary_phi_akp99}d, we estimate the scattering optical depth $\tau = \kappa \rho H$, where $\kappa$ ($= 0.38$ cm$^2$ g$^{-1}$) is the electron scattering opacity, and $H$ is the disk half-thickness. We find that disk continues to remain optically thin ($\tau < 1$) in the post-shock regime, which eventually indicates that the emergent high energy radiations can easily escape from the PSC. Next, in Fig. \ref{fig:vary_phi_akp99}e, we show the variation of radial ($b^r$; thin lines) and toroidal ($b^\phi$; thick lines) magnetic fields as a function of $r$. We notice that $b^r$ increases monotonically with decreasing $r$ following $b^r \propto r^{-2}$ profile. We also observe that in the pre-shock regime, the toroidal field follows the self-similar profile as $b^\phi \propto r^{-5/4}$, and it is amplified at $r_{\rm sh}$ just to maintain the continuity of radial flux ($\Phi_-=\Phi_+$) across the shock front. Thereafter, $b^\phi$ continues to follow a steeper power-law as $b^\phi \propto r^{-7/2}$. With this, the toroidal magnetic field reaches up to $\sim 10^{7-9}$ Gauss near the horizon for the chosen accretion solutions, where magnetic activities are strongest. However, the radial magnetic field limits itself within $\sim 10^{5-6}$ Gauss. An equivalent assessment of magnetic activity is illustrated with the variation of plasma$-\beta$ $(= p_{\rm gas}/p_{\rm mag})$ in Fig. \ref{fig:vary_phi_akp99}f. As the flow starts accreting towards the BH, gas pressure ($p_{\rm gas}$) initially dominates over the magnetic pressure ($p_{\rm mag}$), which enhances $\beta$ values. Indeed, $\beta$ decreases at PSC as $b^\phi$ jumps up higher and it yields magnetically stronger PSC, although flow remains thermal pressure dominated ($\beta > 1$). We further notice that the magnetization $\sigma~(=b^2/\rho$) varies with $r$ as shown in Fig. \ref{fig:vary_phi_akp99}g and $\sigma$ becomes roughly $\sim 100$ times higher near the horizon as compared to the outer edge value. Finally, we present the variation of magnetic pressure corresponding to the radial and toroidal components, and find that toroidal magnetic pressure dominates the disk magneto-hydrodynamics as $b^\phi b_\phi > b_r b^r $ all throughout the disk including near horizon domain (see Fig. \ref{fig:vary_phi_akp99}h).

\begin{figure}
    \centering
    \includegraphics[width=\columnwidth]{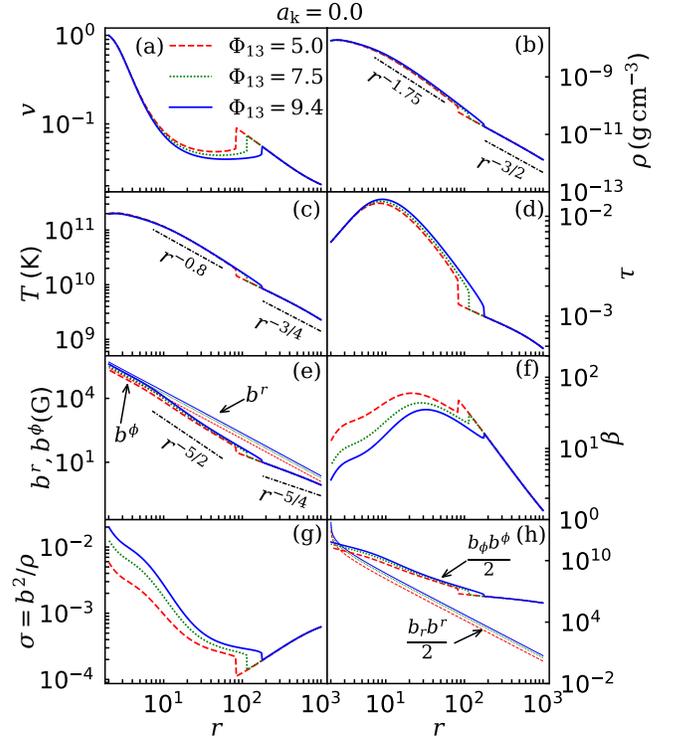}
    \caption{Same as Fig. \ref{fig:vary_phi_akp99}, but flow variables correspond to shocked solutions presented in Fig. \ref{fig:1}a. Here, dashed (red), dotted (green) and solid (blue) curves denote results for $\Phi_{\rm 13}= 5.0$, $7.0$ and $9.4$, respectively. See the text for the details.} 
    \label{fig:vary_phi_ak0}
\end{figure}

It is useful to examine the properties of the primitive flow variables focusing on a non-rotating black hole ($a_{\rm k}=0.0$), as illustrated in Fig. \ref{fig:vary_phi_ak0}. The model parameters remain consistent with those in Fig. \ref{fig:vary_phi_akp99}, except for $\mathcal{L}=3.15$. The radial variations of the flow variables exhibit qualitative similarity with the results obtained for solutions around rapidly rotating BH of spin $a_{\rm k}=0.99$. Nevertheless, as the shock transition tends to occur at relatively larger radii around a non-rotating BH ($a_{\rm k}=0.0$), the compression at the post-shock flow weakens. As a result, the density ($\rho$), temperature ($T$), and magnetic fields near the black hole decrease. This results in shallower fitting profiles of the post-shock flow variables in panel (b) $\rho \propto r^{-1.75}$, (c) $T \propto r^{-0.8}$, and (e) $b^\phi \propto r^{-5/2}$, compared to the results obtained for $a_{\rm k}=0.99$. Notably, the pre-shock variables exhibit the same radial dependency as observed in Fig. \ref{fig:vary_phi_akp99}, regardless of the spin of BH. Finally, it is observed that the radial component of magnetic pressure ($b^rb_r/2$) exceeds the toroidal component ($b^\phi b_\phi/2$) near the horizon (as shown in Fig. \ref{fig:vary_phi_ak0}h), which contrasts with the rotating black hole case (as shown in Fig. \ref{fig:vary_phi_akp99}h).

\subsection{Shock properties}

It is intriguing to investigate the effect of magnetic fields on the shock properties, namely shock location ($r_{\rm sh}$), compression ratio ($R$), and shock strength ($\Psi$) as the spectral properties of BH often rely on these quantities \cite[]{Chakrabarti-Titurchuk1995,Nandi-etal2012,Nandi-etal2018}. Towards this, we examine how shock properties change with radial magnetic flux ($\Phi_{13}$) and iso-rotation parameter ($F_{15}$) for GRMHD flows accreting on to rapidly rotating BH of spin $a_{\rm k}=0.99$. 

\begin{figure}
    \centering
    \includegraphics[width=\columnwidth]{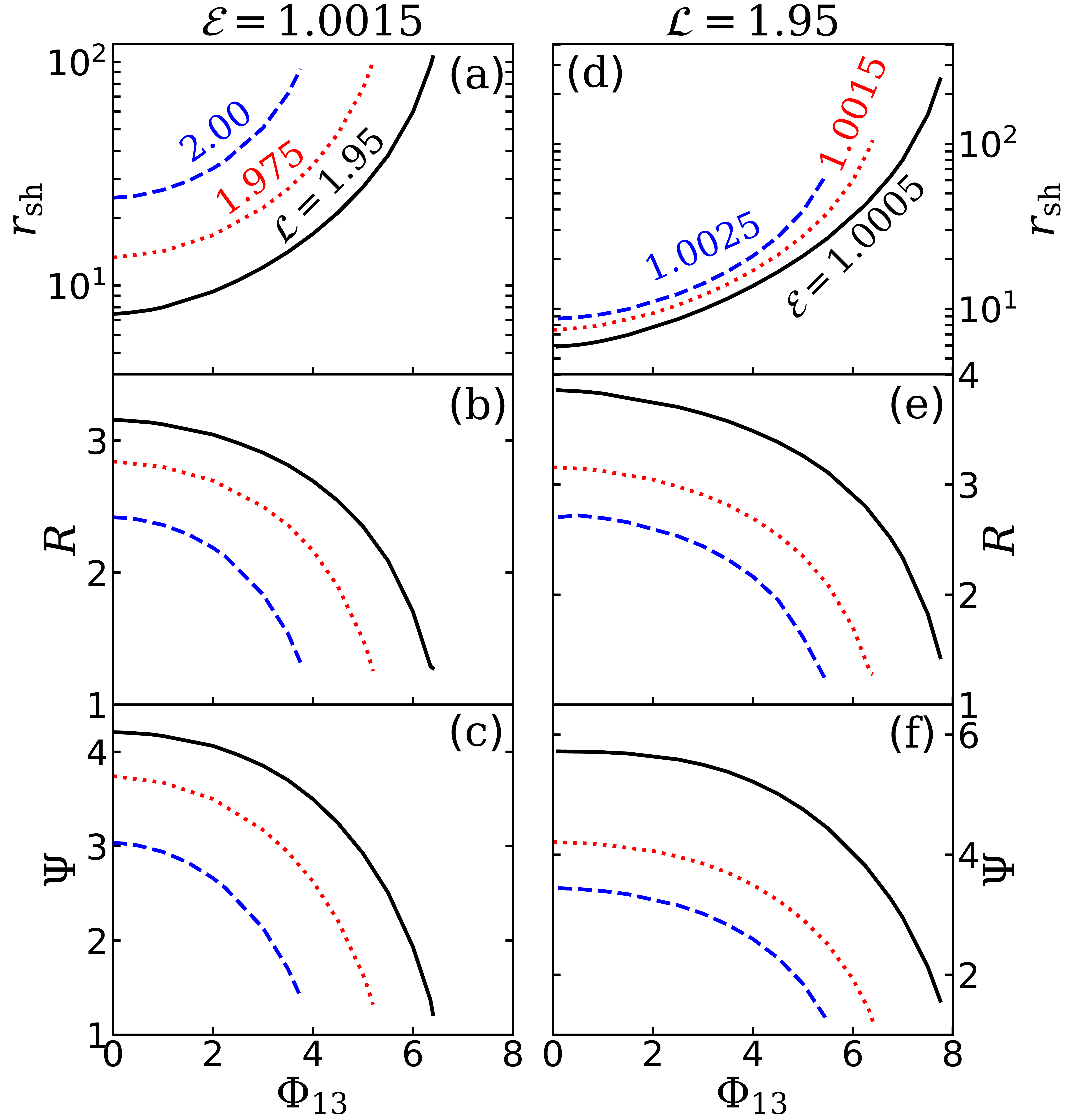}
    \caption{Variation of (a, d) shock location $r_{\rm sh}$, (b, e) compression ratio $R$, and (c, f) shock strength $\Psi$ with $\Phi_{13}$. In left panels, angular momentum fluxes is varied as ${\cal L}=1.95$, $1.975$ and $2.00$, keeping ${\cal E}= 1.0015$ fixed. Similarly, in right panels, we vary energy as ${\cal E}=1.0005$, $1.0015$ and $1.0025$ for fixed ${\cal L}=1.95$. Remaining model parameters are set as $a_{\rm k}=0.99$ and $F_{15}=5.0$. See the text for the details.}
    \label{fig:Shock_rs_R_S_Phi}
\end{figure}

\begin{figure}
    \centering
    \includegraphics[width=\columnwidth]{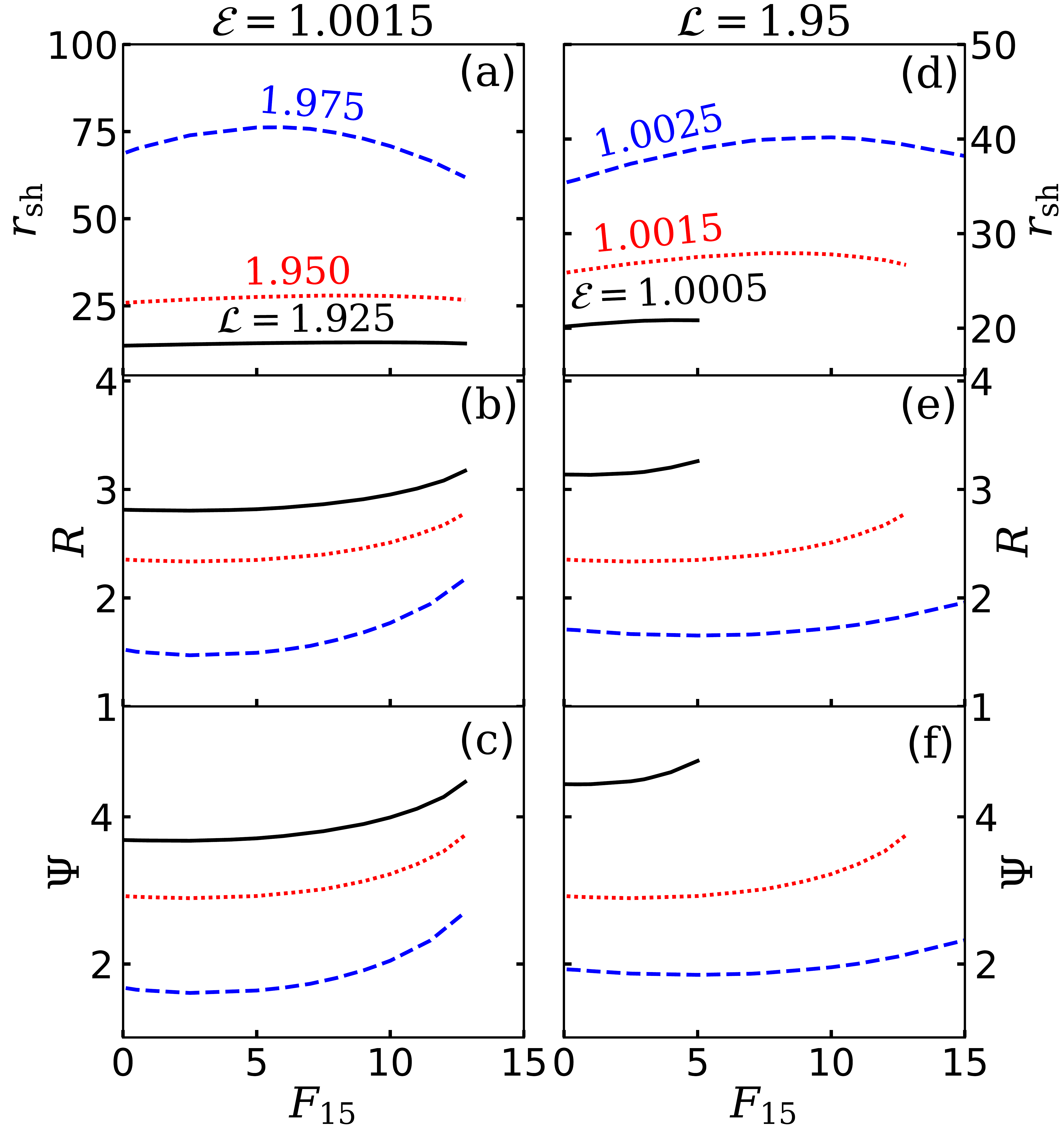}
    \caption{Variation of (a, d) shock location $r_{\rm sh}$, (b, e) compression ratio $R$, and (c, f) shock strength $\Psi$ with $F_{15}$. In left panels, angular momentum fluxes is varied as ${\cal L}=1.925$, $1.950$ and $1.975$, keeping ${\cal E}= 1.0015$ fixed. Similarly, in right panels, we vary energy as ${\cal E}=1.0005$, $1.0015$ and $1.0025$ for fixed ${\cal L}=1.95$. Remaining model parameters are set as $a_{\rm k}=0.99$ and $\Phi_{13}=5.0$. See the text for the details.}
    \label{fig:Shock_rs_R_S_F}
\end{figure}

In Fig. \ref{fig:Shock_rs_R_S_Phi}a, we depict the variation of shock location ($r_{\rm sh}$) as a function of radial magnetic flux ($\Phi_{13}$) for different values of angular momentum ($\cal L$). Here, we choose ${\cal E}=1.0015$ and $F_{15}=5.0$. Solid, dotted and dashed curves represent results for ${\cal L}=1.95$, $1.975$ and $2.0$, respectively.
We observe that for the chosen set of model parameters ($\mathcal{E,L},F_{15}$), standing shocks form for minimum radial flux limit $\Phi_{13}^{\rm min}=0$. Moreover, we find that for a fixed $\cal L$, standing shocks continue to form at larger radii as $\Phi_{13}$ is increased until it reaches a critical limit ($\Phi_{13}^{\rm cri}$). Beyond $\Phi_{13}^{\rm cri}$, shock disappears as shock conditions are not satisfied. Indeed, $\Phi_{13}^{\rm cri}$ does not possess universal value as it depends on the other model parameters. Further, we notice that for fixed $\Phi_{13}$, shocks form at larger radii as $\cal L$ is increased. This evidently indicates that standing shocks in GRMHD flows seems to be centrifugally supported. Indeed, it is useful to study the density profile of the GRMHD flow as the emitted radiations directly depends on it. Meanwhile, we find that convergent GRMHD flow experiences density compression across the shock front $r_{\rm sh}$ (see Figs. \ref{fig:vary_phi_akp99}b, \ref{fig:vary_phi_ak0}b). Accordingly, we compute the compression ratio $R$ defined as the ratio of surface mass density ($\Sigma=\rho H$) of post-shock and pre-shock flow and depict it in Fig. \ref{fig:Shock_rs_R_S_Phi}b as function of $\Phi_{13}$ for the same set of model parameters as in Fig. \ref{fig:Shock_rs_R_S_Phi}a. For a fixed $\cal L$, $R$ decreases with higher $\Phi_{13}$. This happens because enhanced $\Phi_{13}$ increases the magnetic pressure inside the disk and hence, shock front is pushed outward resulting the weakening of density compression due to expansion of PSC size. Similarly, for a given $\Phi_{13}$, GRMHD flow with higher $\cal L$ experiences less compression at the PSC as increased centrifugal pressure counteracts the inward motion of the flow. Overall, we observe that strong shock ($R \rightarrow 4$) exists for smaller $\Phi_{13}$, whereas shock tends to become weak ($R\rightarrow 1$) for larger $\Phi_{13}$. We further compute shock strength ($\Psi$) that accounts the temperature jump across the shock front. The shock strength is defined as the ratio of pre-shock to post-shock Mach numbers as $\Psi=\frac{v_-/C_{{\rm s}_-}}{v_+/C_{{\rm s}_+}}$, and we plot $\Psi$ in Fig. \ref{fig:Shock_rs_R_S_Phi}c as a function of $\Phi_{13}$ for identical model parameters as in Fig. \ref{fig:Shock_rs_R_S_Phi}a. We find that for a given $\cal L$, $\Psi$ is stronger when $\Phi_{13}$ is smaller and vice versa. Moreover, we observe that $\Psi$ exhibits a similar trend to that of the compression ratio ($R$). 

In Fig. \ref{fig:Shock_rs_R_S_Phi}d, we show the variation $r_{\rm sh}$ with $\Phi_{13}$ for different values of flow energy ($\cal E$). Here, we choose $\mathcal{L}=1.95$ and $F_{15}=5.0$. Solid, dotted and dashed curves denote results corresponding to ${\cal E}=1.0005$, $1.0015$ and $1.0025$, respectively. We find that for a fixed $\cal E$, shock settles down at larger radii as $\Phi_{13}$ is increased. Indefinite increase of radial magnetic flux is not possible as there exists a cut-off value of $\Phi_{13}$ for which shock conditions are not satisfied. Furthermore, we calculate $R$ and $S$ as in Fig. \ref{fig:Shock_rs_R_S_Phi}b-c and observe that both quantities are decreased when $\Phi_{13}$ is increased.

In Fig. \ref{fig:Shock_rs_R_S_F}, we depict the comparison of shock properties as a function of the iso-rotation parameter ($F_{15}$). In panels (a-c) of Fig. \ref{fig:Shock_rs_R_S_F}, we plot $r_{\rm sh}$, $R$ and $\Psi$ for different values of $\cal L$. Here, we choose the model parameters as $\mathcal{E}=1.0015$, $\Phi_{13}=5.0$ and $a_{\rm k}=0.99$. The solid, dotted and dashed curves denote results for ${\cal L}=1.925$, $1.950$ and $1.975$, respectively. On the contrary, in Fig. \ref{fig:Shock_rs_R_S_F}d-f, we present the results of $r_{\rm sh}$, $R$ and $\Psi$ for different $\cal E$, where solid, dotted and dashed are for 
${\cal E}= 1.0005$, $1.0015$ and $1.0025$, respectively. The model parameters are chosen as ${\cal L}=1.95$, $\Phi_{13}=5.0$ and $a_{\rm k}=0.99$. In both scenarios, we notice that the shock location remains nearly unaffected due to the increase in $F_{15}$. Consequently, both compression ratio ($R$) and shock strength ($\Psi$) exhibit negligible variation with $F_{15}$ as well. Because of this, now onwards, we refrain examining the influence of $F_{15}$ on the shock properties unless stated otherwise.

\begin{figure}
	\centering
    \includegraphics[width=\columnwidth]{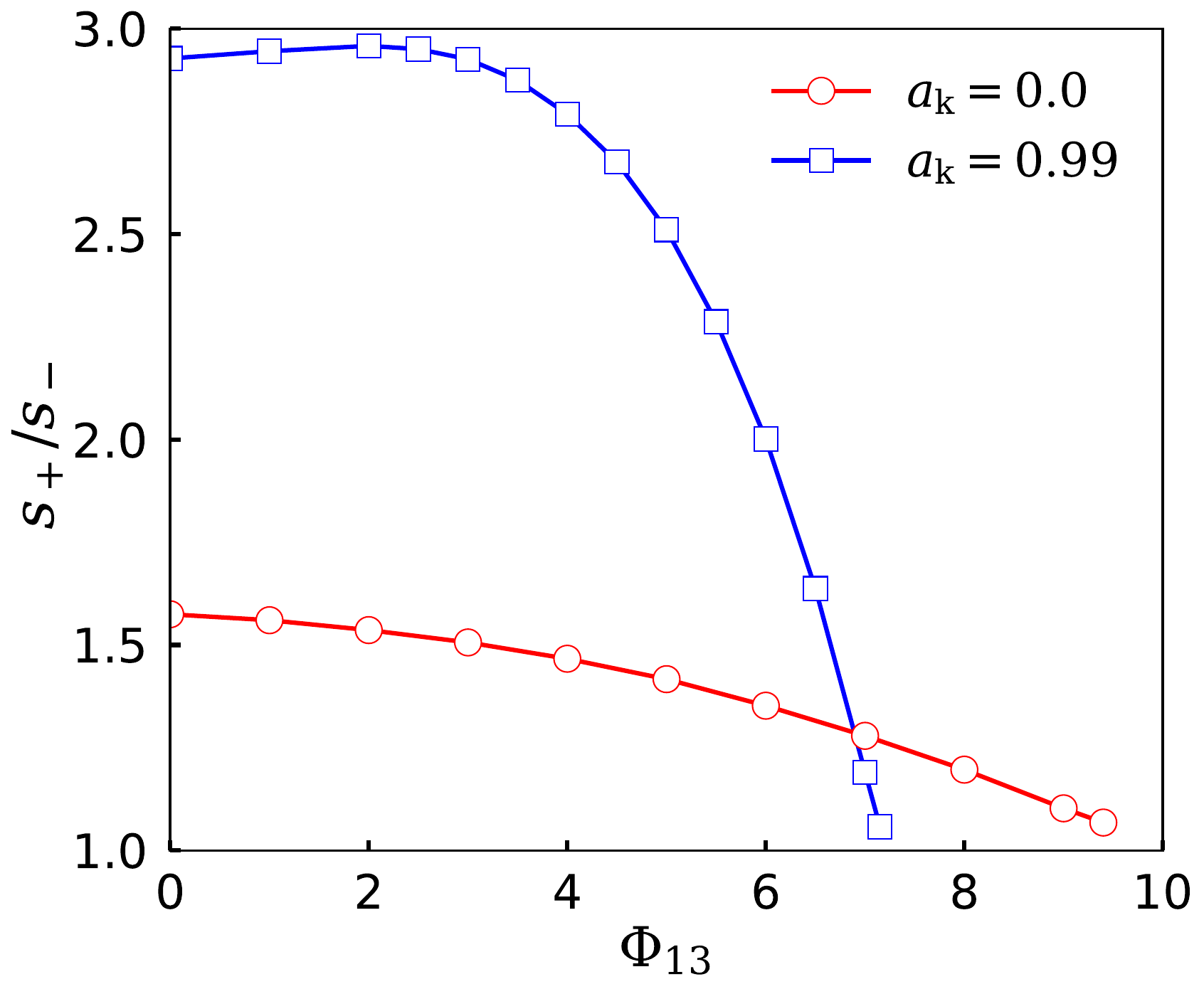}
    \caption{Plot of ratio of post-shock ($s_{+}$) to pre-shock ($s_{-}$) entropy functions across the shock front with $\Phi_{13}$. Open circles and open squares joined with solid lines denote results obtained for non-rotating ($a_{\rm k}=0.0$) and rapidly rotating ($a_{\rm k}=0.99$) BHs. Here, we choose $\mathcal{L}=3.15$ for $a_{\rm k}=0.0$ and $\mathcal{L}=1.95$ for $a_{\rm k}=0.99$, while other model parameters are kept fixed as $\mathcal{E}=1.001$ and $F_{15}=7.5$. See the text for the details.
    }
    \label{fig:ENT}
\end{figure}

As previously noted, shock-induced global GRMHD solutions are favoured over shock-free solutions due to their elevated entropy content. However, the role of the magnetic fields in contributing to the flow entropy is not well understood. To address this, we calculate the ratio of entropy functions measured immediately after ($s_{+}$) and before ($s_{-}$) the shock transition. The obtained results are depicted in Fig. \ref{fig:ENT}, where we plot the variation of $s_{+}/s_{-}$ with $\Phi_{13}$ for flows with fixed energy $\mathcal{E}=1.001$ and iso-rotation parameter $F_{15}=7.5$. In the figure, open circles joined with solid lines represent results corresponding to the shocked accretion solutions around a non-rotating BH with $a_{\rm k}=0.0$ and $\mathcal{L}=3.15$. Similarly, open squares joined with solid lines are for rapidly rotating BH with $a_{\rm k}=0.99$, and $\mathcal{L}=1.95$. We observe that for a chosen set of model parameters, $s_{+}/s_{-}$ is maximum for $\Phi_{13}=0.0$ irrespective to $a_{\rm k}$ values, and it generally decreases with the increase of $\Phi_{13}$. When $\Phi_{13}$ reaches a critical limit, the ratio $s_{+}/s_{-}$ approaches unity, indicating that shock ceases to exist as shock conditions are not favourable. Indeed, it's worth noting that this critical limit of $\Phi_{13}$ isn't universal, as it varies depending on the other model parameters.

\subsection{Parameter space for standing shock}

\begin{figure}
    \centering
    \includegraphics[width=\columnwidth]{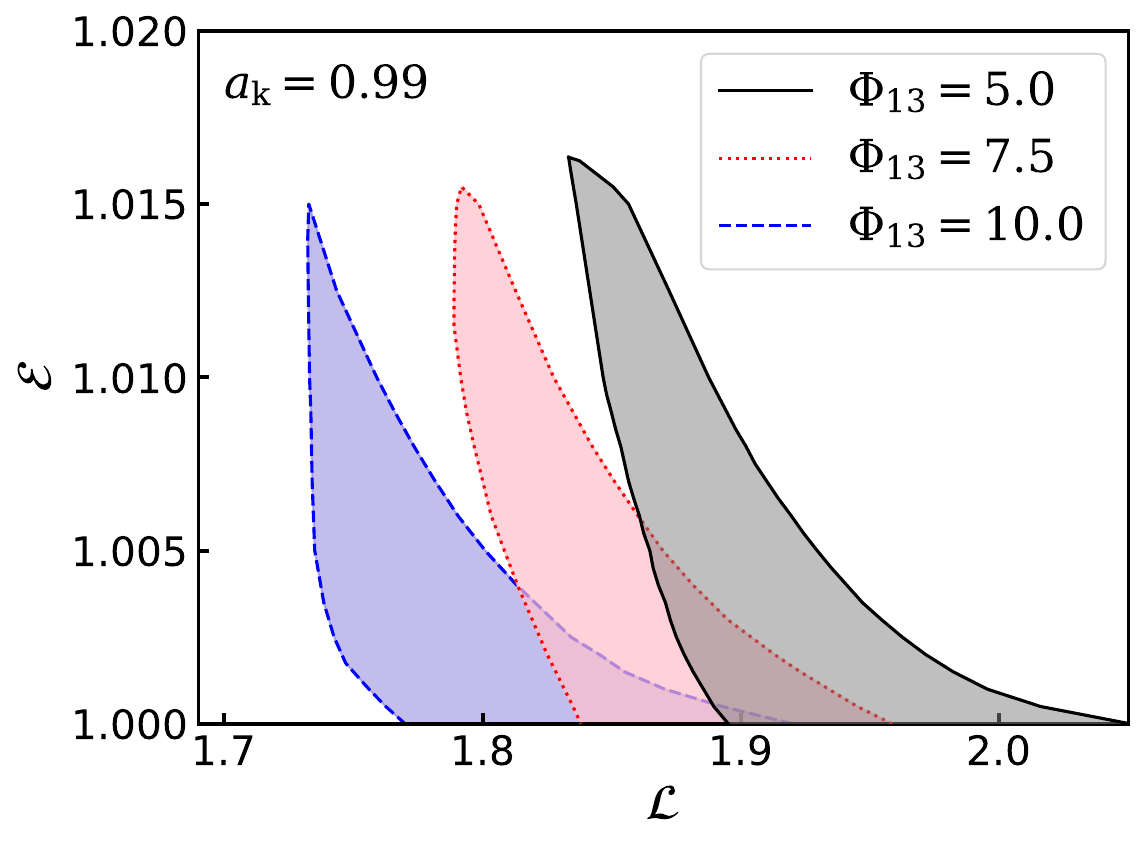}
    \caption{Plot of shock parameter space in $\mathcal{L}-\mathcal{E}$ plane for different radial magnetic fluxes ($\Phi_{13}$) around a rotating BH. Here, we choose $a_{\rm k}=0.99$ and $F_{15}=5.0$. Effective area bounded by solid (black), dotted (red) and dashed (blue) curves correspond to $\Phi_{13}=5.0, 7.5$ and $10.0$, respectively.  
    See the text for the details.}
    \label{fig:Paraspace-1}
\end{figure}

\begin{figure}
 \centering
 \includegraphics[width=\columnwidth]{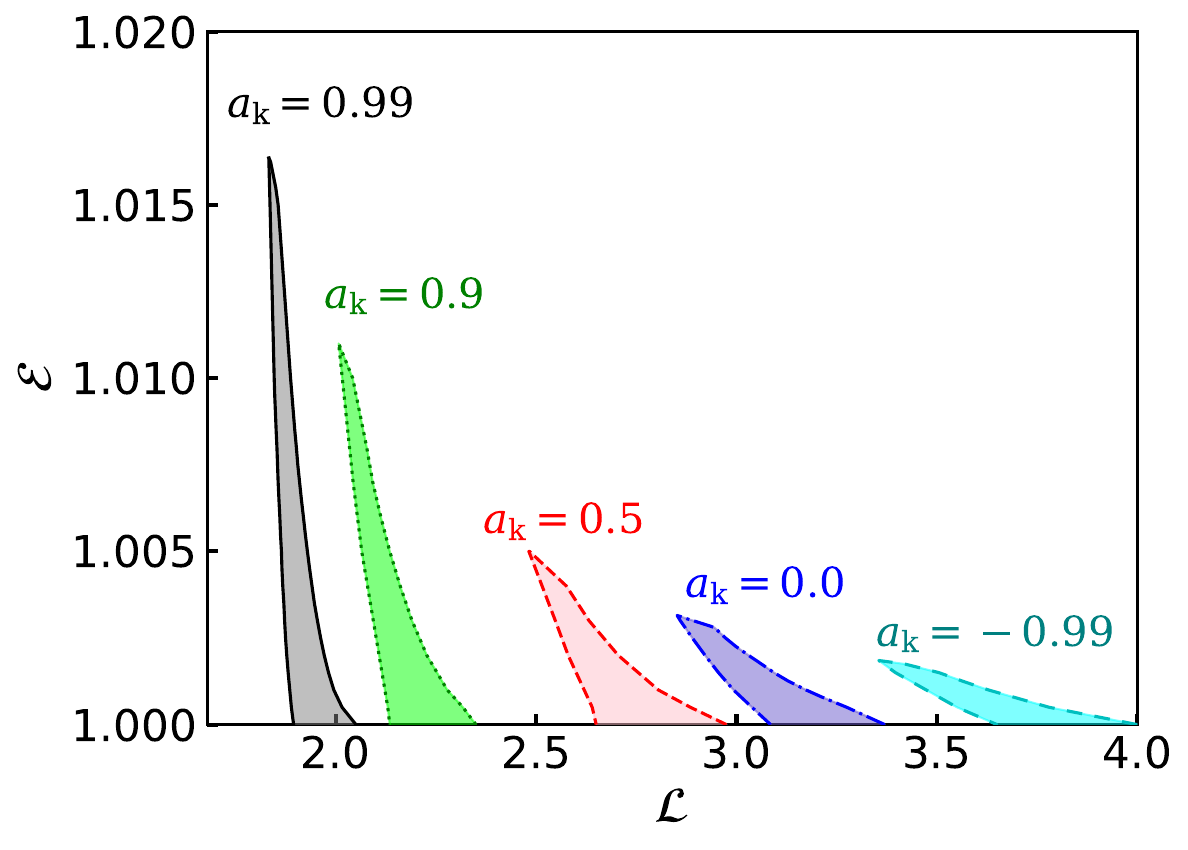}
 \caption{Modification of shock parameter space in $\mathcal{L}-\mathcal{E}$ plane with BH spin varied as $a_{\rm k}=0.99, 0.9, 0.5, 0.0$ and $-0.99$ (left to right). Here, we fix $\Phi_{13}=5.0$ and $F_{15}=5.0$. See the text for the details.}
 \label{fig:Paraspace-2}
\end{figure}

It has already been indicated that shock-induced GRMHD accretion solutions are not isolated solutions, instead these solutions continue to exist for a wide range of model parameters, namely $\cal E$, $\cal L$, $\Phi_{13}$, $F_{15}$, and $a_{\rm k}$. Hence, it is useful to identify the ranges of model parameters that admit shocked accretion solutions. Towards this, in Fig. \ref{fig:Paraspace-1}, we separate the effective domain of the parameter space in ${\cal L}-{\cal E}$ plane and examine the modification of the parameter space for different $\Phi_{13}$, where spin of the BH and $F_{15}$ are kept fixed as $a_{\rm k}=0.99$ and $5.0$. The region enclosed by solid (black), dotted (red), and dashed (blue) curves are for $\Phi_{13}=5.0, 7.5$, and $10.0$, respectively. We observe that parameter space shifts towards the lower angular momentum ($\mathcal{L}$) domain as $\Phi_{13}$ is increased. This happens because the increase of $\Phi_{13}$ effectively enhances the angular momentum transport outwards \cite[]{Mitra-etal2022} leading to the reduction of $\cal L$, which allows shock transition for higher $\Phi_{13}$. Similarly, in Fig. \ref{fig:Paraspace-2}, we present the modification of the parameter space for different $a_{\rm k}$. Here, we choose $\Phi_{13}=5.0$ and $F_{15}=5.0$. We observe that the effective region bounded by different line style are obtained for $a_{\rm k}=0.99$, $0.9$, $0.5$, $0.0$, and $-0.99$ (from left to right). We again observe that parameter space is shifted to the lower angular momentum side as the BH spin is increased \cite[]{Aktar-etal2015,Dihingia-etal2019}. This is not surprising because relatively low angular momentum flow experiences shock transition around BH of higher $a_{\rm k}$ \cite[]{Das-Chakrabarti2008}. Overall, we stress that both $a_{\rm k}$ and $\Phi_{13}$ play pivotal role in determining the shock parameter space of GRMHD flow.

\begin{figure}
    \centering
    \includegraphics[width=\columnwidth]{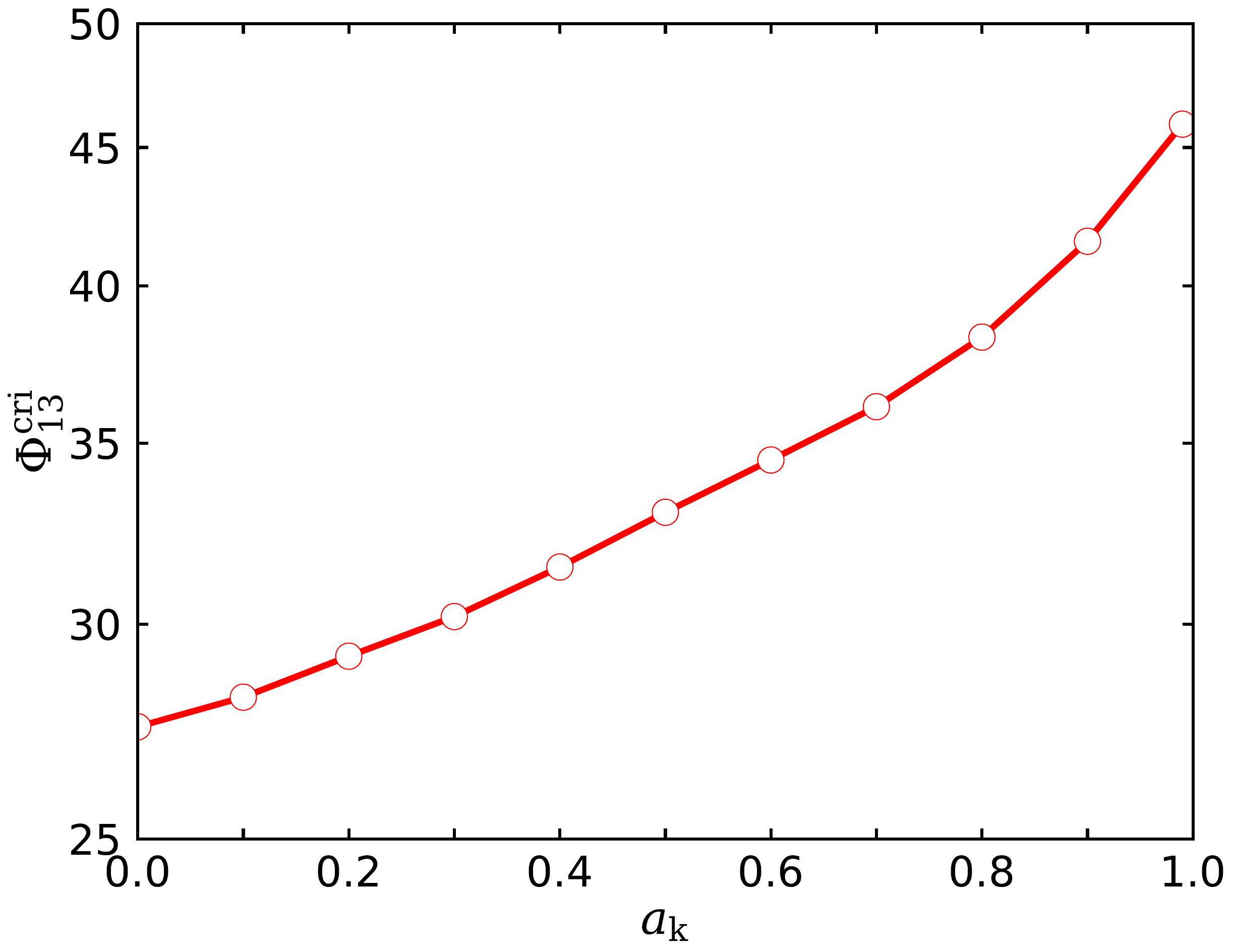} 
    \caption{Variation of maximum radial magnetic flux ($\Phi^{\rm max}_{13}$ as a function of BH spin ($a_{\rm k}$) for shocked accretion solutions. Here, we choose $F_{\rm 15}=5.0$, and $(\mathcal{E,L})$ are chosen freely. See text for details.}
    \label{fig:phi_max_ak}
\end{figure}

Subsequently, we investigate the critical radial magnetic flux ($\Phi^{\rm cri}_{13}$) necessary to render global GRMHD accretion solutions around BHs harbouring shocks. In doing so, we calculate $\Phi^{\rm cri}_{13}$ for various values of BH spin ($a_{\rm k}$), while keeping the iso-rotation parameter and accretion rate fixed at $F_{15}=5.0$ and ${\dot m}=0.001$, respectively, and the energy ($\mathcal{E}$) and angular momentum ($\mathcal{L}$) of flow are allowed to vary freely. The obtained results are depicted in Fig. \ref{fig:phi_max_ak}, where open circles connected by solid lines represent the variation of $\Phi^{\rm cri}_{13}$ with $a_{\rm k}$. We observe that shock-induced GRMHD accretion solutions exist for a wide range of $\Phi_{13}$. In particular, we find that for $a_{\rm k}=0.0$, $\Phi^{\rm cri}_{13}=27.5$, and $\Phi^{\rm cri}_{13}$ increases with the increase of $a_{\rm k}$, where $\Phi^{\rm cri}_{13}=45.9$ for $a_{\rm k}=0.99$. This finding evidently indicates that GRMHD accretion flow with stronger magnetic fields continue to sustain shocks around rapidly rotating black hole and vice versa. What is more is that by analysing the disk magnetic flux, one can speculate whether the magnetized accretion flow is consistent with the results of either Magnetically Arrested Disk (MAD) or Standard and Normal Evolution (SANE) state \cite[]{Narayan-etal2012, Chatterjee-Narayan2022}). To assess this, we calculate the upper limit of magnetic flux entering the ergosphere as $\int_{0}^{2\pi} -\sqrt{-g} \,\,^*{F^{rt}} d\phi= 3.23 \times 10^{18} \Phi^{\rm cri}_{13}$ G cm$^2$, and find that the GRMHD accretion flows under consideration remain significantly below the MAD limit $\sim 10^{21}$ G cm$^2$, as indicated by \cite{Sadowski2016b}, for a black hole with a mass of $10M\odot$ and a spin parameter within the range $0 \le a_{\rm k} < 0.99$.

\section{Conclusions}
\label{Conclusion}

In this work, we study the global structure of shock-induced, magnetized, advective accretion flow around rotating BHs. In doing this, we solve the general relativistic magnetohydrodynamics (GRMHD) equations that govern the flow motion in the steady state and obtain comprehensive solutions for global trans-magnetosonic accretion flow around weakly ($a_{\rm k}\rightarrow0$) as well as rapidly ($a_{\rm k}=0.99$) rotating BHs. We observe that depending on the model parameters, namely $\mathcal{E}$, $\mathcal{L}$, $\Phi_{13}$, $F_{15}$ and $a_{\rm k}$, GRMHD flow often possess multiple critical points ($r_{\rm in}$ and $r_{\rm out}$). It is noteworthy that the accretion solutions simultaneously passing through both $r_{\rm in}$ and $r_{\rm out}$ are of special importance as they may harbour shock waves and shock-induced accretion solutions are potentially promising in explaining the spectro-temporal properties of BH sources \cite[]{Chakrabarti-Titurchuk1995,Nandi-etal2012,Das-etal2021}. With this, we summarize our present findings below.

\begin{itemize}

\item We obtain shock-induced global GRMHD accretion solutions around rotating BHs for the first time to the best of our knowledge. We observe that irrespective to the BH spin ($a_{\rm k}$), shock forms further out as the radial magnetic flux ($\Phi_{13}$) is increased for flows with fixed model parameters ($\mathcal E$, $\mathcal L$, $F_{15}$). We further notice that the entropy of the shocked accretion solution always remains higher than the shock free solution, confirming that shocked solutions are thermodynamically preferred (see Fig. \ref{fig:1}).

\item We observe that magnetic fields play an important role in regulating the structure of the shocked GRMHD accretion flow around BHs (see Fig. \ref{fig:vary_phi_akp99}-\ref{fig:vary_phi_ak0}). In general, the accretion flow largely remains gas pressure dominated ($\beta > 10$) throughout the disk, expect at the inner part of the disk ($r < 10 r_g$). Moreover, we find that in the pre-shock regime, toroidal field follows self-similar radial profile as $b^\phi \propto r^{-5/4}$ \cite[]{Yuan-Narayan2014}, however, it becomes stepper in the post-shock flow in order to maintain the continuity of radial flux balance across the shock front (see \S 3). The toridal field becomes more intense around the rotating BHs due to the effect of frame-dragging, and for a $10M_\odot$ BH of spin $a_{\rm k}=0.99$, it becomes very strong as $b^\phi \sim 10^{7-9}$ G near the horizon. On the contrary, we find $b^\phi \sim 10^{5-6}$ G in the region close to horizon of a non-rotating ($a_{\rm k}=0.0$) BH.

\item We also examine the best fit power-law profile of density ($\rho$) and temperature ($T$) in both pre- and post-shock regimes around weakly rotating ($a_{\rm k} \rightarrow 0.0$) as well as rapidly rotating ($a_{\rm k}=0.99$) black holes. We ascertain that in the pre-shock regime, the density and temperature profiles of GRMHD flow remain unaffected due to BH spin and are obtained as $\rho \propto r^{-3/2}$ and $T\propto r^{-3/4}$, respectively. This findings are in agreement with the results of \cite{Narayan-Yi1994,Yuan-Narayan2014}. However, due to shock compression, both $\rho$ and $T$ follow steeper power-law profile as $\rho \propto r^{-2}$ and $T\propto r^{-1}$ for $a_{\rm k}=0.99$, and $\rho \propto r^{-1.75}$ and $T\propto r^{-0.8}$ for $a_{\rm k}=0.00$ (see Fig. \ref{fig:vary_phi_akp99}-\ref{fig:vary_phi_ak0}).

\item Convergent GRMHD shocked accretion flow yields hot and dense post-shock flow (see Fig. \ref{fig:vary_phi_akp99}-\ref{fig:vary_phi_ak0}) resembling a post-shock corona (PSC), containing hot electrons. When soft photons from the pre-shock flow interact with the PSC, they undergo inverse Comptonization, producing hard X-ray radiations commonly observed from black hole X-ray binaries \cite[]{Chakrabarti-Titurchuk1995,Nandi-etal2012,Iyer-etal2015,Nandi-etal2018}. Since the PSC characteristics ($i.e.$, its size, density and temperature) are determined by the shock properties, and magnetic fields ($\Phi_{13}$ and $F_{15}$) controls $r_{\rm sh}$, $R$ and $\Psi$ (see Figs. \ref{fig:Shock_rs_R_S_Phi}-\ref{fig:Shock_rs_R_S_F}), it is therefore evident that magnetic fields play a crucial role in determining the spectral properties of black holes.

\item Moreover, we find that the shock-induced GRMHD accretion solutions are not isolated solutions as solutions of this kind continue to exist for a wide range of model parameters, namely $\mathcal{E}$, $\mathcal{L}$, $\Phi_{13}$, $F_{15}$ and $a_{\rm k}$. To ascertain this, we separate the parameter space in $\mathcal{L}-\mathcal{E}$ plane for different $\Phi_{13}$ that admits GRMHD shock solutions around rapidly rotating ($a_{\rm k}=0.99$) BHs (see Fig. \ref{fig:Paraspace-1}). Moreover, we examine the modification of the shock parameter space (in $\mathcal{L}-\mathcal{E}$ plane) for various BH spin ($a_{\rm k}$) values and observe that GRMHD accretion flows exhibit shocks within the spin range $-0.99 < a_{\rm k} < 0.99$ (see Fig. \ref{fig:Paraspace-2}).

\item We calculate the critical radial magnetic flux ($\Phi^{\rm cri}_{13}$), representing the threshold beyond which shocks cease to exist in GRMHD accretion flow around BHs. We observe a strong dependence of $\Phi^{\rm cri}_{13}$ on the spin of the black hole ($a_{\rm k}$), with $\Phi^{\rm cri}_{13}$ being higher for $a_{\rm k}=0.99$ compared to $a_{\rm k}=0.0$ (see Fig. \ref{fig:phi_max_ak}). We also observe that shock-induced GRMHD accretion flow under consideration remains restricted below the MAD threshold.

\end{itemize}

Finally, we indicate the limitations of the present work as it is carried out based on assumptions. We neglect the polar component of the magnetic fields ($b^\theta$) considering the fact that disk is confined around the equatorial plane. Indeed, $b^\theta$ is expected to play crucial role in launching jets and outflows \cite[][and references therein]{Dihingia-etal2021}. We ignore the effect of radiative coolings although their presence are relevant.  Moreover, we consider single temperature fluid neglecting two temperature descriptions of ions and electrons. We also work out adopting the ideal MHD limit, ignoring resistive MHD approach. Indeed, the implementation of these complexities exceeds the scope of the present paper, and we plan to take them up in future endeavors.

\section*{Data Availability}
 
The data underlying this article will be available with reasonable request.

\section*{Acknowledgements}

Authors thank the anonymous reviewer for constructive comments and useful suggestions that help to improve the quality of the manuscript. SM acknowledges Prime Minister's Research Fellowship (PMRF), Government of India for financial support. SM is indebted to Indu K. Dihingia and Debaprasad Maity for illuminating discussions. The work of SD is supported by the Science and Engineering Research Board (SERB), India, through grant MTR/2020/000331.

\bibliography{reference}{}
\bibliographystyle{aasjournal}

\appendix
\section{Derivation of Wind Equations}

The radial momentum equation, angular momentum equation and energy equation are obtained as,

\begin{equation}
R_0 + R_v \frac{dv}{dr} + R_\Theta \frac{d\Theta}{dr} + R_\lambda \frac{d\lambda}{dr} =0,
\end{equation}
\begin{equation}
		\mathcal{L}_0 + \mathcal{L}_v \frac{dv}{dr} + \mathcal{L}_\Theta \frac{d\Theta}{dr} + \mathcal{L}_\lambda \frac{d\lambda}{dr} =0,
	\end{equation}
\begin{equation}
		\mathcal{E}_0 + \mathcal{E}_v \frac{dv}{dr} + \mathcal{E}_\Theta \frac{d\Theta}{dr} + \mathcal{E}_\lambda \frac{d\lambda}{dr} =0,
\end{equation}
and the coefficients in equations (A1-A3) are given below. In order to express these coefficients, we begin with the derivative of the four-velocities $u_{\mu}'=\frac{du^\mu}{dr}$, where all the Greek symbols correspond to the following coordinates $\mu \equiv (t,r,\theta,\phi)$, 

$$u_{\mu}'= u_{\mu_0} + u_{\mu_v} v' + u_{\mu_\lambda} \lambda',	u^{\mu'}= u^{\mu}_0 + u^{\mu}_v v' + u^{\mu}_\lambda \lambda',$$
$$u^{\mu}_0=\frac{\partial u^\mu}{\partial r}, u^\mu_v=\frac{\partial u^\mu}{\partial v}, u^\mu_\lambda=\frac{\partial u^\mu}{\partial \lambda}, u_{\mu_0}=\frac{\partial u_\mu}{\partial r}, u_{\mu_v}=\frac{\partial u_\mu}{\partial v}, u_{\mu_\lambda}=\frac{\partial u_\mu}{\partial \lambda}.$$
With the following definitions of $b^r$, $b^\phi$ (see Eq. 12) and $b^t$,
$$
b^r = - \frac{\gamma_\phi^2 (\Phi + F \lambda)}{u^t r^2 (v^2 - 1)},\;\;\;
b^\phi = \frac{F v^2 - \gamma_\phi^2 (F + \Phi \Omega)}{u^r r^2 (v^2 - 1)}, \,\,\, b^t = -\frac{u_r}{u_t} b^r + \lambda b^\phi,$$
we can write,
$$ b^{\mu'}= b^{\mu}_0 + b^{\mu}_v v' + b^{\mu}_\lambda \lambda', {\rm where}\,\,  b^{\mu}_0=\frac{\partial b^\mu}{\partial r}, b^\mu_v=\frac{\partial b^\mu}{\partial v}, b^\mu_\lambda=\frac{\partial b^\mu}{\partial \lambda}.$$ 
In this way, the derivative of the square of the magnetic field $\mathcal{B}=b_\mu b^\mu$ is expressed as,
$$ \mathcal{B}'= \mathcal{B}_0 + \mathcal{B}_v v' + \mathcal{B}_\lambda \lambda', \,\,\, \mathcal{B}_0=\frac{\partial \mathcal{B}}{\partial r}, \mathcal{B}_v=\frac{\partial \mathcal{B}}{\partial v}, \mathcal{B}_\lambda=\frac{\partial \mathcal{B}}{\partial \lambda}.$$ 
Similarly, 
$$h_{\rm tot}' = h_{\rm t}^0 + h_{\rm t}^v v' + h_{\rm t}^\Theta \Theta' + h_{\rm t}^\lambda \lambda', h_{\rm t}^0 = \frac{\partial{h_{\rm tot}}}{\partial r},  h_{\rm t}^v = \frac{\partial{h_{\rm tot}}}{\partial v}, h_{\rm t}^\Theta = \frac{\partial{h_{\rm tot}}}{\partial \Theta}, h_{\rm t}^\lambda = \frac{\partial{h_{\rm tot}}}{\partial \lambda},$$ and
$$\mathcal{F}'=\frac{d\mathcal{F}}{dr}=\mathcal{F}_1 + \mathcal{F}_2 \frac{d\lambda}{dr},
\mathcal{F}_1 = \frac{\partial\mathcal{F}}{\partial r}, \mathcal{F}_2 = \frac{\partial\mathcal{F}}{\partial \lambda}, \Delta' = \frac{d\Delta}{dr}.$$

With these definitions, we express the coefficients of the equations (A1-A3) as follows,
$$ R_0 =(R_a + \mathcal{A} R_1)/\rho h_{\rm tot}, \, \mathcal{A}=(g^{rr} + u^r u^r), \, R_1 =   \frac{\mathcal{B}_0}{2}  - \frac{3 \Theta \rho }{ r \tau} + \frac{\mathcal{F}_1 \Theta \rho}{\tau \mathcal{F}} - \frac
{\Theta \rho \Delta'}{\tau \Delta},$$ $$R_a = - b^r b^r_0 (2 + g_{rr} u^r u^r) + h_{\rm tot} u^r u^r_0 \rho - b^r u^r (u_t b^t_0 + u_\phi b^\phi_0) + R_b, \, R_b = -g^{rr} b^{r^2} (1 - \frac{1}{2} g_{rr} u^{r^2}) g_{rr}' + R_c,$$
$$ R_c = \frac{\rho h_{\rm tot}}{2} g^{rr} u^{r^2} g_{rr}' + \frac{g^{rr} b^{t^2}}{2} g_{tt}' + g^{rr} g_{t\phi}' b^t b^\phi + \frac{g^{rr}}{2} b^{\phi^2} g_{\phi\phi}' - \frac{g^{\theta\theta}}{2} b^{r^2} g_{\theta\theta}' - \mathcal{S}_1 (b^{r^2} + 2 b^r u^r b^t u_t) + R_d, $$ 
$$ R_d =  - 2 b^r u^r b^t u_\phi \mathcal{S}_2 - 2 b^r u^r b^\phi u_t \mathcal{S}_2  -\mathcal{S}_3 g_{rr} u^{r^2} b^t + \mathcal{S}_4 h_{\rm tot} u^t \rho - g_{rr} u^{r^2} b^\phi \mathcal{S}_5 + h_{\rm tot} \rho u^\phi \mathcal{S}_6 - \mathcal{S}_7 (b^{r^2} + 2 b^r u^r b^\phi u_\phi ), $$ 
$$\mathcal{S}_1=\frac{1}{2} (g^{tt} g_{tt}' + g^{t\phi} g_{t\phi}'),\, \mathcal{S}_2=\frac{1}{2} (g^{tt} g_{t\phi}' + g^{t\phi} g_{\phi\phi}'), \, \mathcal{S}_3 = -\frac{g^{rr}}{2}(b^t g_{tt}' + b^\phi g_{t\phi}'), \, \mathcal{S}_4 = -\frac{g^{rr}}{2} (u^t g_{tt}' + u^\phi g_{t\phi}'),$$
$$\mathcal{S}_5 = -\frac{g^{rr}}{2} (b^t g_{t\phi}' + b^\phi g_{\phi\phi}'), \mathcal{S}_6 = -\frac{g^{rr}}{2} (u^t g_{t\phi}' + u^\phi g_{\phi\phi}'), \mathcal{S}_7 = \frac{1}{2} (g^{t\phi} g_{t\phi}' + g^{\phi\phi} g_{\phi\phi}').$$

$$ R_v =  (R_e + \mathcal{A} R_2)/\rho h_{\rm tot}, \, R_2 =   \frac{\mathcal{B}_v}{2}  - \frac{2 \Theta \rho (1+v^2 \gamma_v^2)}{ \tau \mathcal{F}}, R_e = - b^r b^r_v (2 + g_{rr} u^r u^r) + h_{\rm tot} u^r u^r_v \rho - b^r u^r (u_t b^t_v + u_\phi b^\phi_v),$$
$$ R_\Theta = \frac{1}{\tau h_{\rm tot}},\,  R_\lambda = (R_f + \mathcal{A} R_3)/\rho h_{\rm tot}, \, R_3 = \frac{\mathcal{B}_\lambda}{2}  + \frac{\mathcal{F}_2 \Theta \rho }{ \tau \mathcal{F}}, R_f = - b^r b^r_\lambda ( 2 + g_{rr} u^r u^r) - b^r u^r (u_t b^t_\lambda + u_\phi b^\phi_\lambda).$$

$$ \mathcal{L}_0 = h_t^0 u_\phi + h_{\rm tot} u_{\phi_0} - \frac{g_{t\phi} b^r b^t_0 + b^r_0 b_\phi + g_{\phi\phi} b^r b^\phi_0 - b^r b^t g_{t\phi}' + b^r b^\phi g_{\phi\phi}' }{u^r \rho} - \frac{3 b^r b_\phi}{2 r u^r \rho} + \frac{\mathcal{F}_1 b^r b_\phi}{2 \mathcal{F} u^r \rho}  - \frac{b^r b_\phi \Delta'}{2 u^r \Delta \rho} + \frac{b^r u^r_0 b_\phi}{u^{r^2} \rho},$$

$$ \mathcal{L}_v = h_t^v u_\phi + h_{\rm tot} u_{\phi_v} - \frac{ g_{t\phi} b^r b^t_v + b^r_v b_\phi + g_{\phi\phi} b^r b^\phi_v }{u^r \rho} + \frac{b^r u^r_v b_\phi}{u^{r^2} \rho} - \frac{b^r (1+v^2 \gamma_v^2) b_\phi}{u^r v \rho},$$

$$ \mathcal{L}_\Theta  = h_t^\Theta u_\phi - \frac{b^r b_\phi}{2 u^r \Theta \rho },\,  \mathcal{L}_\lambda = h_t^\lambda u_\phi + h_{\rm tot} u_{\phi_\lambda} - \frac{g_{t\phi} b^r b^t_\lambda}{u^r \rho} + \frac{\mathcal{F}_2 b^r b_\phi}{2 \mathcal{F} u^r \rho} - \frac{b^r_\lambda b_\phi}{u^r \rho} - \frac{g_{\phi\phi} b^r b^\phi_\lambda}{u^r \rho}.$$

$$ \mathcal{E}_0 = -h_t^0 u_t - h_{\rm tot} u_{t_0} + \frac{g_{tt} b^r b^t_0 + b^r_0 b_t + g_{t\phi} b^r b^\phi_0 - b^r b^\phi g_{t\phi}' + b^r b^t g_{tt}' }{u^r \rho} + \frac{3 b^r b_t}{2 r u^r \rho} - \frac{\mathcal{F}_1 b^r b_t}{2 \mathcal{F} u^r \rho}  - \frac{b^r b_t \Delta'}{2 u^r \Delta \rho} - \frac{b^r u^r_0 b_t}{u^{r^2} \rho},$$

$$ \mathcal{E}_v = -h_t^v u_t - h_{\rm tot} u_{t_v} + \frac{ g_{tt} b^r b^t_v + b^r_v b_t + g_{t\phi} b^r b^\phi_v }{u^r \rho} - \frac{b^r u^r_v b_t}{u^{r^2} \rho} + \frac{b^r (1+v^2 \gamma_v^2) b_t}{u^r v \rho},$$

$$ \mathcal{E}_\Theta  = -h_t^\Theta u_t + \frac{b^r b_t}{2 u^r \Theta \rho },\,  \mathcal{E}_\lambda = -h_t^\lambda u_t - h_{\rm tot} u_{t_\lambda} + \frac{g_{tt} b^r b^t_\lambda}{u^r \rho} - \frac{\mathcal{F}_2 b^r b_t}{2 \mathcal{F} u^r \rho} + \frac{b^r_\lambda b_t}{u^r \rho} + \frac{g_{t\phi} b^r b^\phi_\lambda}{u^r \rho}.$$

\section{Expressions of Numerator and Denominator}

As mentioned previously, we express the wind equation (18) as,
\begin{equation}
\frac{dv}{dr} = \frac{\mathcal{N}(r,a_{\rm k},v,\Theta,\lambda,\Phi, F)}{\mathcal{D}(r,a_{\rm k},v,\Theta,\lambda,\Phi, F)}.
\end{equation}
Here, the numerator ($\cal N$) is given by,
\begin{equation}
\mathcal{N} = - R_0 - R_\Theta \Theta_{11} - R_\lambda \lambda_{11},
\end{equation}
and the denominator ($\cal D$) is given by,
\begin{equation}
\mathcal{D}= R_v + R_\Theta \Theta_{12} + R_\lambda \lambda_{12},
\end{equation}
where 
$$ \Theta_{11} = \frac{\mathcal{E}_\lambda \mathcal{L}_0 - \mathcal{E}_0 \mathcal{L}_\lambda}{-\mathcal{E}_\lambda \mathcal{L}_\Theta + \mathcal{E}_\Theta \mathcal{L}_\lambda}, \,\, \Theta_{12} = \frac{\mathcal{E}_\lambda \mathcal{L}_v - \mathcal{E}_v \mathcal{L}_\lambda}{-\mathcal{E}_\lambda \mathcal{L}_\Theta + \mathcal{E}_\Theta \mathcal{L}_\lambda}, \,\, \lambda_{11}=\frac{-\mathcal{E}_\Theta \mathcal{L}_0 + \mathcal{E}_0 \mathcal{L}_\Theta}{-\mathcal{E}_\lambda \mathcal{L}_\Theta + \mathcal{E}_\Theta \mathcal{L}_\lambda}, \,\, \lambda_{12}=\frac{-\mathcal{E}_\Theta \mathcal{L}_v + \mathcal{E}_v \mathcal{L}_\Theta}{-\mathcal{E}_\lambda \mathcal{L}_\Theta + \mathcal{E}_\Theta \mathcal{L}_\lambda}.$$

\end{document}